%% file: main.tex
\documentclass[sigconf]{acmart}

\usepackage{hyperref}
\usepackage{lipsum}
\usepackage{setspace}
\usepackage{booktabs, tabularx}
\usepackage{threeparttable}
\usepackage{array, xspace, subfigure, url, xcolor, amsmath, bm, array,amsfonts, balance,multirow, color, enumitem,kantlipsum, float, multicol}

\usepackage{tcolorbox}

\usepackage{graphicx}  
\usepackage{float}  

\definecolor{darkblue}{rgb}{0,0,0.5}
\definecolor{darkred}{rgb}{0.5,0,0}
\definecolor{darkgreen}{rgb}{0,0.5,0}
\definecolor{balanceblue}{rgb}{0.25,0,0.25}
\definecolor{balance}{rgb}{0.25,0.25,1.0}
\definecolor{extensionA}{rgb}{0.25,0.5,1.0}
\definecolor{extensionB}{rgb}{1.0,0.25,0.25}
\definecolor{extensionC}{rgb}{0,0.5,0.5}
\definecolor{extensionD}{rgb}{1.0,0.5,0.5}
\definecolor{extensionE}{rgb}{0.1,0.1,0.5}
\definecolor{comment}{rgb}{0,0,0.5}
\hypersetup{
	colorlinks=true,
	urlcolor=darkblue,
	linkcolor=darkblue,
	citecolor=darkblue
}

\newcommand{\init}{\textcolor{extensionE}{\textsf{init}}\xspace}
\newcommand{\vrf}{\textcolor{extensionD}{\textsf{VerifyAggregation}}}
\newcommand{\cvk}{\textcolor{extensionD}{\textsf{ConstructVK}}}
\newcommand{\verify}{\textcolor{extensionD}{\textsf{Verify}}}
\newcommand{\verifyphase}{\textcolor{extensionD}{\textsf{VerifyPhase}}}
\newcommand{\register}{\textcolor{extensionD}{\textsf{CommitModel}}}

\newcommand{\add}{\textcolor{extensionD}{\textsf{Addition}}}
\newcommand{\smul}{\textcolor{extensionD}{\textsf{ScalarMul}}}
\newcommand{\negate}{\textcolor{extensionD}{\textsf{Negate}}}
\newcommand{\pair}{\textcolor{extensionD}{\textsf{PairingCheck}}}
\newcommand{\newepoch}{\textcolor{extensionD}{\textsf{NewEpoch}}}
\newcommand{\deposit}{\textcolor{extensionD}{\textsf{Deposit}}}
\newcommand{\prepare}{\textcolor{extensionD}{\textsf{Prepare}}}
\newcommand{\preparephase}{\textcolor{extensionD}{\textsf{PreparePhase}}}

\newcommand{\vdf}{\textcolor{extensionD}{\textsf{VDF}}}
\newcommand{\readepoch}{\textcolor{extensionD}{\textsf{ReadEpoch}}}
\newcommand{\claim}{\textcolor{extensionD}{\textsf{ClaimReward}}}

\newcommand{\commit}{\textcolor{extensionD}{\textsf{CommitInputs}}}
\newcommand{\len}{\textcolor{extensionD}{\textsf{Len}}}

\newcommand{\onlyowner}{\textcolor{extensionE}{\textsf{onlyDA}}\xspace}
\newcommand{\payable}{\textcolor{extensionE}{\textsf{payable}}\xspace}
\newcommand{\constructor}{\textcolor{extensionE}{\textsf{constructor}}\xspace}
\newcommand{\modifier}{\textcolor{extensionE}{\textsf{modifier}}\xspace}
\newcommand{\public}{\textcolor{extensionE}{\textsf{public}}\xspace}
\newcommand{\private}{\textcolor{extensionE}{\textsf{private}}\xspace}

\newcommand{\epoch}{\textsf{epoch}}
\newcommand{\train}{\textsf{Training}\xspace}
\newcommand{\req}{\textcolor{balance}{\textsf{require}}\xspace}
\newcommand{\struct}{\textcolor{balance}{\textsf{struct}}\xspace}

\newcommand{\gone}{\textcolor{extensionC}{\textsf{G1Point}}\xspace}
\newcommand{\gtwo}{\textcolor{extensionC}{\textsf{G2Point}}\xspace}
\newcommand{\uuint}{\textcolor{extensionC}{\textsf{uint}}\xspace}
\newcommand{\aaddr}{\textcolor{extensionC}{\textsf{address}}\xspace}
\newcommand{\bbool}{\textcolor{extensionC}{\textsf{bool}}\xspace}
\newcommand{\sstring}{\textcolor{extensionC}{\textsf{string}}\xspace}

\newcommand{\mmapping}{\textcolor{extensionC}{\textsf{mapping}}\xspace}
\newcommand{\tru}{\textcolor{extensionC}{\textsf{true}}\xspace}
\newcommand{\fal}{\textcolor{extensionC}{\textsf{false}}\xspace}
\newcommand{\nnull}{\textcolor{extensionC}{\textsf{null}}\xspace}
\newcommand{\msg}{\textcolor{extensionC}{\textsf{msg}}\xspace}
\newcommand{\timestamp}{\textcolor{extensionC}{\textsf{block.timestamp}}\xspace}
\newcommand{\abort}{\textcolor{extensionC}{\textsf{abort}}\xspace}
\newcommand{\emit}{\textcolor{extensionC}{\textsf{emit}}\xspace}

\newcommand{\return}{\textsf{\textbf{return}}\xspace}
\newcommand{\func}{\textsf{\textbf{Function}}\xspace}
\newcommand{\mycomment}[1]{\textcolor{darkblue}{\footnotesize{\emph{\textsf{#1}}}}}

\newcommand{\ie}{\emph{i.e.,}\xspace}
\newcommand{\eg}{\emph{e.g.,}\xspace}
\newcommand{\sys}{\textsf{martFL}\xspace}
\newcommand{\first}{\textsf{(i)}\xspace}
\newcommand{\second}{\textsf{(ii)}\xspace}
\newcommand{\third}{\textsf{(iii)}\xspace}
\newcommand{\fourth}{\textsf{(iv)}\xspace}

\newcommand{\dacquirer}{\textsf{DA}\xspace}
\newcommand{\dprovider}{\textsf{DP}\xspace}
\newcommand{\dproviders}{\textsf{DPs}\xspace}

\newcommand{\bco}{\textbf{:}\xspace}
\newcommand\revision[1]{{\color{black} #1}}

\makeatletter
\newif\if@restonecol
\makeatother

\usepackage[linesnumbered,ruled,vlined]{algorithm2e}
\usepackage[noend]{algpseudocode}

\SetCommentSty{mycommfont}

\renewcommand\footnotetextcopyrightpermission[1]{} 
\usepackage[skip=0.0pt]{caption}
\usepackage[misc]{ifsym}

\newcommand{\paraspace}{\vspace{0.02in}}
\newcommand{\parab}[1]{\paraspace\noindent{\bf #1}}

\newif\iftechreport 
\techreporttrue

\copyrightyear{2023} 
\acmYear{2023} 
\setcopyright{acmlicensed}\acmConference[CCS '23]{Proceedings of the 2023 ACM SIGSAC Conference on Computer and Communications Security}{November 26--30, 2023}{Copenhagen, Denmark}
\acmBooktitle{Proceedings of the 2023 ACM SIGSAC Conference on Computer and Communications Security (CCS '23), November 26--30, 2023, Copenhagen, Denmark}
\acmPrice{15.00}
\acmDOI{10.1145/3576915.3623134}
\acmISBN{979-8-4007-0050-7/23/11}



\begin{document}
\title[\sys: A Robust and Verifiable FL for Data Marketplace]{\sys: Enabling Utility-Driven Data Marketplace with a Robust and Verifiable Federated Learning Architecture}
\titlenote{This is a longer version of the paper originally published in ACM CCS 2023~\cite{martfl-ccs}.}

\author{Qi Li}
\affiliation{%
\institution{Tsinghua University \& Zhongguancun Laboratory}
}
\email{li-q20@mails.tsinghua.edu.cn}

\author{Zhuotao Liu}
\affiliation{%
\institution{Tsinghua University \& Zhongguancun Laboratory}
}
\email{zhuotaoliu@tsinghua.edu.cn}
\authornote{Zhuotao Liu is the corresponding author.}

\author{Qi Li}
\affiliation{%
\institution{Tsinghua University \& Zhongguancun Laboratory}
}
\email{qli01@tsinghua.edu.cn}

\author{Ke Xu}
\affiliation{%
\institution{Tsinghua University \& Zhongguancun Laboratory}
}
\email{xuke@tsinghua.edu.cn}

\input{0.abs}

\maketitle

\input{1.intro}

\input{2.background}

\input{3.framework}

\input{4.model_evaluation}
\input{5.verifiable_transaction}

\input{6.evaluation}

\input{discussion}
\input{7.related}

\input{8.conclusion}


\balance
\bibliographystyle{acm}
\bibliography{ref} 

\iftechreport
\newpage
\input{appendix}
\fi

\end{document}

%% file: 0.abs.tex
\begin{abstract}
The development of machine learning models requires a large amount of training data. Data marketplace is a critical platform to trade high-quality and private-domain data that is not publicly available on the Internet. However, as data privacy becomes increasingly important, directly exchanging raw data becomes inappropriate. 
Federated Learning (FL) is a distributed machine learning paradigm that exchanges data utilities (in form of local models or gradients) among multiple parties without directly sharing the original data. However, we recognize several key challenges in applying existing FL architectures to construct a data marketplace. \first In existing FL architectures, the Data Acquirer (\dacquirer) cannot privately assess the quality of local models submitted by different Data Providers (\dproviders) prior to trading; \second The model aggregation protocols in existing FL designs cannot effectively exclude malicious \dproviders without ``overfitting'' to the \dacquirer's (possibly biased) root dataset; 
\third Prior FL designs lack a proper billing mechanism to enforce the \dacquirer to fairly allocate the reward according to contributions made by different \dproviders. 
To address above challenges, we propose \sys, the first federated learning architecture that is specifically designed to enable a secure utility-driven data marketplace. At a high level, \sys is empowered by two innovative designs: \first a quality-aware model aggregation protocol that allows the \dacquirer to properly exclude local-quality or even poisonous local models from the aggregation, even if the \dacquirer's root dataset is biased; 
\second a verifiable data transaction protocol that enables the \dacquirer to prove, both succinctly and in zero-knowledge, that it has faithfully aggregated these local models according to the weights that the \dacquirer has committed to. This enables the \dproviders to unambiguously claim the rewards proportional to their weights/contributions. 
We implement a prototype of \sys and evaluate it extensively over various tasks. The results show that \sys can improve the model accuracy by up to 25\% while saving up to 64\% data acquisition cost. 

\end{abstract}

%% file: 1.intro.tex
\section{Introduction}\label{sec:introduction}

Artificial Intelligence (AI) continues to shape many aspects of our lives. However, the development of AI models requires a large amount of high-quality training data. However, collecting data, especially the private-domain data that is not publicly available on the Internet, is challenging. 
The community proposed the concept of data marketplace~\cite{fernandez2020data,koutsos2020agora, krishnamachari2018i3, niu2018achieving} to address this problem. In a data marketplace (such as the International Data Spaces Association~\cite{internationaldataspacesassociation}), organizations can access high-quality data owned by other organizations that is specific to their needs. 
However, as data privacy becomes increasingly important, directly trading raw data could be inappropriate or even prohibited by laws (\eg GDPR \cite{voigt2017eu}, PIPL \cite{tan2021china}). 
This implies a fundamental paradigm shift from trading raw data to only trading data utilities without raw data exchange. 

\begin{figure*}[!t]
\centering
\includegraphics[width=2\columnwidth]{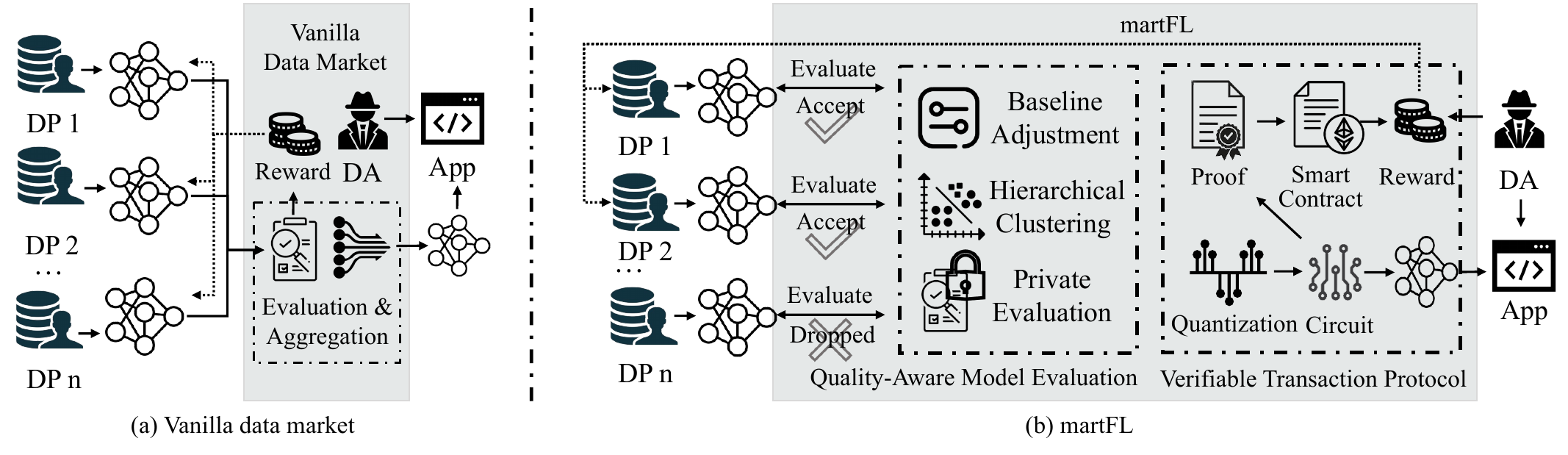}
\caption{The architecture comparison between the vanilla FL and \sys.}
\label{fig:arch}
\end{figure*}

Federated Learning (FL)~\cite{mcmahan2017communication}  is a machine learning paradigm that enables multiple parties to train a global model on their own data without sharing the data with each other. This is achieved by having each party train a local model on their own data and then sending the updates to a central server. The central server then aggregates the updates from all of the parties to create a global model.
This makes FL a promising paradigm for a utility-driven marketplace because organizations can buy and sell data without having to share the underlying data.
Viewing FL as a primitive, we could construct a strawman data marketplace as shown in Figure~\ref{fig:arch}(a). 
In this diagram, the aggregation server in FL serves as the \emph{data acquirer} (\dacquirer) that initiates the FL task. The FL clients serve as the \emph{data providers} (\dproviders) to participate in the FL task by providing local model updates trained on their own data. 
The \dacquirer evaluates the local models submitted by different \dproviders to purchase high-quality local models. Eventually, the \dacquirer aggregates local models to update the global model, based on which it may initiate another iteration. 


However, we recognize three major challenges in applying the vanilla FL to construct a secure data marketplace. 
First, the model aggregation protocol in vanilla FL does not allow the \dacquirer to evaluate the data quality of local model updates before obtaining the updates from \dproviders. This raises a dilemma in data trading: the \dproviders are unwilling to give away their local updates before receiving rewards, while the \dacquirer prefers to evaluate the updates first before purchasing them. 

Second, the model aggregation protocol of the vanilla FL is subject to various attacks, such as~\cite{bagdasaryan2020backdoor, lin2019free, zhu2020advanced, fung2020limitations}. Prior art on mitigating these issues can be roughly categorized into client-driven approaches and server-driven approaches. 
The client-driven approaches~\cite{xu2020reputation, blanchard2017machine, yin2018byzantine} improve aggregation robustness by smoothing the local update updates based on their statistics (\eg median or average). 
The server-driven approaches~\cite{lyu2020collaborative, cao2021fltrust} instead rely on the \dacquirer to lead the aggregation process. They assume that the \dacquirer possesses a high-quality \emph{root dataset} based on which it can calibrate the local model updates submitted by the \dproviders. 
As a result, prior works exhibit a fundamental tradeoff between inclusiveness and robustness: the client-driven approaches can potentially include more local model updates for training, yet they stake robustness on the ``honest majority'' (which might be incorrect in the data trading scenario), while the server-driven approaches are more resilient against malicious \dproviders, yet they sacrifice inclusiveness by ``overfitting'' to the \dacquirer's existing dataset (which could be biased). 

Third, existing model aggregation protocols lack the required \emph{verifiability} to enable fair billing. Specifically, in FL, the local models receiving higher aggregation weights have more impacts on the final model. Thus, the aggregation weights essentially quantify the values (or utilities) provided by the local model updates. 
Prior art (\eg Omnilytics~\cite{liang2021omnilytics}, FPPDL~\cite{lyu2020towards}) tries to achieve fair billing by directly executing \emph{the entire model aggregation process} on blockchains, which significantly limits the design space of the aggregation algorithms (see analysis in \S~\ref{subsec:background:FL}). 
Thus, the third challenge in designing a fair utility-driven data marketplace is to ensure that the \dacquirer faithfully distributes rewards among the \dproviders according to their actual aggregation weights. This also ensures that the \dacquirer only pays for its desired model updates, rather than blindly purchasing arbitrary updates, which greatly reduces model acquisition cost (see evaluation results in \S~\ref{subsec:evaluation_results}).

To address these challenges, we present \sys, a novel FL architecture that enables robust and verifiable local model aggregation in a utility-driven data marketplace. 
\sys is powered by two innovative designs.  First, \sys designs a two-phased protocol that first privately evaluates all local model updates submitted by \dproviders based on a baseline to remove outliers (\ie local-quality updates) and then dynamically adjusts the evaluation baseline to incorporate the high-quality updates. 
Therefore, our quality-aware model aggregation protocol eliminates the fundamental tradeoff between inclusiveness and robustness, by indiscriminately evaluating the complete set of \dproviders and meanwhile avoiding overfitting to the (possibly biased) root dataset owned by \dacquirer. 

Second, \sys designs a novel verifiable data transaction protocol that enables the \dacquirer and the selected \dproviders to securely exchange the reward and model updates. 
Our verifiable transaction protocol centers around a proving scheme that allows the \dacquirer to prove, both succinctly and in zero-knowledge, that it faithfully aggregates the model using the committed aggregation weights. Based on the publicly verifiable proof, the \dproviders can unambiguously claim the reward corresponding to their weights.   
Crucially, \sys achieves the fair trading without relying on any online trusted third party to regulate the trading process.



\parab{Contributions.} The main contribution of this paper is the design, implementation and evaluation of \sys, the first FL architecture that simultaneously offers robustness and verifiability to enable a secure utility-driven data marketplace. 
We implement a prototype of \sys in approximately 3750 lines of code and extensively evaluate its accuracy and robustness using two image classification datasets and two text classification datasets. The results show that compared to existing server-driven methods, \sys can improve accuracy by up to 25\% even when the \dacquirer has a biased root dataset, while saving up to 64\% data acquisition cost. In addition, \sys can resist various untargeted attacks, targeted attacks, and Sybil attacks, and achieves the highest accuracy and the lowest attack success rate compared to both server-driven and client-driven methods. We also report the system-level overhead of \sys to demonstrate its feasibility in practice. 


%% file: 2.background.tex
\section{Background and Motivation}\label{sec:bg}


\subsection{Data Marketplace}
The traditional circulation of data trading  mainly relies on data trading platforms (such as International Data Spaces~\cite{internationaldataspacesassociation}, BDEX~\cite{bdex_2022}, Quandl~\cite{demo.quandl.com} and GE Predix~\cite{ge_predix}) that are endorsed by government or industry leaders. 
The research community explored API-based marketplace designs that allow the data acquirers to collect data stream online~\cite{krishnamachari2018i3}. 
Due to the rising importance of data privacy, direct trading of raw data, particularly data associated with personal information~\cite{voigt2017eu,tan2021china}, is subject to significant regulatory burdens in practice. Therefore, it is essential to explore data marketplaces that do not require direct exchange of raw data.

\subsection{Federated Learning and Its Robustness}
\label{subsec:background:FL}
In designing an AI-specific marketplace, Federated Learning (FL)~\cite{mcmahan2017communication} is a promising learning paradigm since it enables collaborative training without directly sharing the raw data. A data marketplace built upon the vanilla FL architecture has three phases: \first global model distribution: the central server (serving as the data acquirer \dacquirer) initializes a global model and distributes it to the clients (serving as the data providers \dproviders); \second local model training: the \dproviders use their local data to train the model and then upload the resulting models (referred to as local models) to the \dacquirer; and \third model aggregation: the \dacquirer aggregates these local models to obtain a new global model. This process repeats for multiple epochs until the \dacquirer obtains a sufficiently accurate global model.

However, the above vanilla FL-driven data marketplace faces several critical challenges. First, FL is known to be vulnerable to various attacks, such as untargeted attack~\cite{bernstein2018signsgd, fang2020local} (\eg the Byzantine clients disrupt the training process by rescaling the sizes of local gradients or randomizing the directions of local gradients), targeted attack~\cite{bagdasaryan2020backdoor} (\eg the Byzantine clients mislead the global model to specifically misclassify certain classes), and Sybil Attack~\cite{fung2020limitations}. 
The community has therefore proposed various robustness FL designs that can be roughly divided into two categories. The client-driven approaches~\cite{fung2020limitations,bernstein2018signsgd,yin2018byzantine,blanchard2017machine,xu2020reputation,ozdayi2021defending} try to exclude malicious local model updates by learning representative statistics from all local models; and the server-driven designs~\cite{lyu2020collaborative,fang2020local,chen2021robust,cao2021fltrust} instead assume that the server owns a trusted root dataset based on which it can calibrate these local models. 
These approaches suffer from a fundamental tradeoff between inclusiveness and robustness, resulting in non-trivial performance degradation (see \S~\ref{subsubsec:key_observation}). 

In addition to the robustness concern, existing FL architectures lack several key features that are essential for data trading. On the one hand, the data acquirer (\dacquirer) cannot assess quality of the local models submitted by different \dproviders prior to trading; on the other hand, the \dproviders are not assured of receiving adequate compensation after submitting their models. 
Several recent approaches (\eg Omnilytics~\cite{liang2021omnilytics}, FPPDL~\cite{lyu2020towards}) try to achieve trading-oriented FL designs by simply executing the \emph{entire model aggregation process} on blockchains, either via general-purpose smart contracts or leveraging specialized block structures. These approaches, however, are fundamentally limited because they force the \dacquirer to make the local model assessment protocol publicly executable on blockchains, preventing the \dacquirer from using proprietary and complex/advanced algorithms. As a result, the aggregation algorithm in FPPDL~\cite{lyu2020towards} is unable to handle malicious \dproviders; and Omnilytics~\cite{liang2021omnilytics} only supports four \dproviders using the simple Secure-Aggregation algorithm~\cite{bonawitz2017practical} with the multi-Krum~\cite{blanchard2017machine} algorithm to remove outliers, while incurring significant gas cost (at least 1000 times more than \sys, as shown in \S~\ref{subsubsec:eval:efficiency}). 
\sys is fundamentally different from these blockchain-based FL approaches because \sys relies on smart contract to verify the correctness of \emph{the offline model assessment and aggregation} performed by the \dacquirer. This enables the \dacquirer to design proprietary and advanced local model evaluation protocols to handle various FL attacks. Additionally, \sys designs a verifiable transaction protocol to ensure the \dacquirer cannot cheat about the reward allocation, even though the \dacquirer uses proprietary model aggregation protocols that are not known to the \dproviders.

\iftechreport
\subsection{Zero-Knowledge Proofs and Verifiable Machine Learning}
\input{zk-related-work}
\else
\subsection{Zero-Knowledge Proofs}
To ensure fair trading, \sys requires the \dacquirer to publicly prove that it has faithfully aggregated local models using the aggregation weights that were committed to before receiving the plaintext local models from the \dproviders. 
This proving process can be formulated as an argument of knowledge for the aggregation protocol, without disclosing these local models to the public. The recent progress in zero-knowledge proof technology, especially the development of zero-knowledge succinct non-interactive arguments of knowledge (zk-SNARK)~\cite{groth2016size,wahby2018doubly,bowe2019recursive,kothapalli2022nova} where the prover only needs present one message (proof) instead of interacting with the verifier~\cite{goldwasser2019knowledge}, has demonstrated the potential to achieve this goal. 
Yet, simply applying existing zk-SNARK constructions~\cite{groth2016size,chiesa2020marlin,bunz2020transparent} to prove the end-to-end training process in FL is challenging. This is because \first the detailed local model evaluation algorithm can be complex and even contains computations over homomorphically encrypted values (see \S~\ref{subsec:model-aggr}); and \second the model sizes are large, for instance, with millions of floating-point parameters or even more. Both of these issues would result in significantly large proof circuits, which are impractical to implement.
\fi

\subsection{Motivation}
To address above challenges, we propose \sys, a secure and verifiable FL architecture specifically designed for utility-driven data marketplaces. \sys advances state-of-the-art in both secure local model aggregation and verifiable data trading. 
In particular, \sys designs a novel quality-aware model evaluation protocol that can indiscriminately and privately assess all the local models submitted by the \dproviders based on a dynamically adjusted baseline. As a result, it can accurately remove malicious local models while avoiding overfitting to the root dataset owned by the \dacquirer, eliminating the tradeoff between inclusiveness and robustness exhibited in prior art. 
Further, \sys proposes an efficient verifiable transaction protocol that enables fair data trading \emph{without the need to prove the entire FL training process}. The key novelty of our approach is that our proving scheme focuses on only proving the critical computation that is necessary and sufficient to ensure fair billing. This results in the proving overhead being independent of both the local model evaluation algorithm and the model size. Given this proof, the \dproviders can unambiguously claim corresponding reward over a smart contract. To the best of our knowledge, this is the first verifiable scheme designed specifically for proving the correctness of model aggregation in FL, without directly placing the entire model aggregation protocol on blockchains.  

\subsection{Assumptions and Threat Model}
We consider Byzantine \dproviders that may submit arbitrary local models. They may launch these aforementioned attacks to disrupt the training process, or try to earn rewards without actual contributions to training (\eg the free-rider attack~\cite{lin2019free}). 
We consider that the \dacquirer is semi-honest, \ie the \dacquirer is protocol-compliant, but motivated to manipulate the reward distribution so as to minimize the cost of collecting data. We assume that the \dacquirer possesses a root dataset. Many well-established robust FL approaches (\eg~\cite{cao2021fltrust,fang2020local}) assumed that DA has a \emph{reliable and unbiased} root or validation dataset to handle malicious \dproviders. In contrast to these approaches, the root dataset assumed in \sys can be \emph{both of poor quality and of limited volume}. For instance, it may contain only half of the labels (\ie the \dacquirer's root dataset exhibits biased distributions), or it may be approximately 1\% of the data held by all \dproviders (see evaluation results in \S~\ref{subsubsec:eval:bias}). 
Therefore, the assumption made about the root dataset in \sys is significantly less restrictive than that made by existing robust FL approaches. This makes \sys suitable for the data trading scenario, in which the \dacquirer, without necessarily possessing a good root dataset, can collect high-quality and high-volume local models from a diverse set of \dproviders.

We assume that the cryptographic primitives and the consensus protocol of the blockchain system used to host the data transaction smart contract in \sys are secure so that the blockchain can have the concept of transaction finality and contract publicity. On Nakamoto consensus based blockchains, finality is achieved by assuming that the probability of blockchain reorganizations drops exponentially as new blocks are appended (\ie the common-prefix property)~\cite{garay2017bitcoin}. On Byzantine tolerance based blockchains, finality is guaranteed by signatures from a quorum of permissioned voting nodes. 
We assume that the blockchain has a public ledger that allows external parties to examine the public state of its deployed smart contracts. 
We assume that the zk-SNARK protocol~\cite{groth2016size} used in our verifiable transaction protocol is sound.

%% file: zk-related-work.tex
To ensure fair trading, \sys requires the \dacquirer to publicly prove that it has faithfully aggregated local models using the aggregation weights that were committed to before receiving the plaintext local models from the \dproviders. 
Formally, this proving process can be formulated as an argument of knowledge for the aggregation protocol, without disclosing these local models to the public. Specifically, an argument of knowledge for an NP relation $\mathcal{R}$ is a protocol between a prover $\mathcal{P}$ and a verifier $\mathcal{V}$. At the end of the protocol, $\mathcal{V}$ is convinced by $\mathcal{P}$ that there exists a witness $w$ such that $(x;w)\in \mathcal{R}$ for some input $x$, without disclosing $w$. 
Let $\mathcal{G}$ denotes the generation algorithm that produces the public parameters $pp$, we have the following formal definition.

\begin{definition}
     A tuple of algorithms $\mathtt{Alg} = (\mathcal{G}, \mathcal{P}, \mathcal{V})$ is a zero-knowledge argument of knowledge for $\mathcal{R}$ if the following holds.
\end{definition}

\begin{itemize}[leftmargin=10pt]
    \item \textbf{Completeness.} For every $pp$ output by $\mathcal{G}(1^\lambda)$, $(x;w)\in \mathcal{R}$,
    $$
    \mathtt{Pr}[\mathcal{V}(pp,x,\mathcal{P}(pp,x,w))= 1] =1
    $$
    \item \textbf{Knowledge Soundness.} For any PPT prover $\mathcal{P^{*}}$, there exists a PPT extractor $\mathcal{E}$ such that $w \leftarrow \mathcal{E}^\mathcal{P^*}$, $\pi \leftarrow \mathcal{P}(pp,x,w)$,
    $$
    \mathtt{Pr}[\left(x;w\right) \notin \mathcal{R} \wedge \mathcal{V}\left(pp,x,\pi\right)= 1 ] \leq \mathtt{negl}\left(\lambda\right)
    $$
    where the extractor $\mathcal{E}^\mathcal{P^*}$ has access to the entire execution, including the randomness of $\mathcal{P^*}$.

    \item \textbf{Zero-knowledge.} There exists a PPT simulator $\mathcal{S}$ such that for any PPT algorithm $\mathcal{V^*}$, it holds that
    $$
    \mathtt{View}(\mathcal{V^*}(pp,x,\pi)) \approx \mathcal{S^\mathcal{V^*}}(x)
    $$
    where $\mathtt{View}(\mathcal{V^*}(pp,x,\pi))$ denotes the view of an honest $\mathcal{V^*}$ in the interaction with $\mathcal{P}$ , $\mathcal{S^\mathcal{V^*}}$ denotes the view generated by $\mathcal{S}$ that given public coin randomness used by $\mathcal{V^*}$, and $\approx$ denotes two perfectly indistinguishable distributions.
\end{itemize}

We say that $\mathtt{Alg} = (\mathcal{G}, \mathcal{P}, \mathcal{V})$ is a succinct argument system if the proof size is $\mathtt{poly}(\log\vert\mathcal{R}\vert,\lambda)$ as well as the running time of $\mathcal{V}$ is $\mathtt{poly}(\vert x \vert,\log\vert\mathcal{R}\vert,\lambda)$, where the $\vert\mathcal{R}\vert$ is the size of the circuit that computes $\mathcal{R}$ as a function of $\mathcal{\lambda}$.

The advances in zero-knowledge proof technology, especially the development of zero-knowledge succinct non-interactive arguments of knowledge (zk-SNARK)~\cite{groth2016size,wahby2018doubly,bunz2018bulletproofs,gabizon2019plonk,bowe2019recursive,kothapalli2022nova} where the prover only needs present one message (proof) instead of interacting with the verifier~\cite{goldwasser2019knowledge}, attracted significant attentions. 
The machine learning community has conducted significant research in applying verifiable computation to achieve privacy and fairness in machine learning services, such as verifiable inference schemes~\cite{ghodsi2017safetynets,zhang2020zero,liu2021zkcnn} that prove the inference results are produced by certain models with claimed accuracies, and verifiable training~\cite{zhao2021veriml} approaches that prove the traceability of training process.

Yet, simply applying existing zk-SNARK constructions~\cite{groth2016size,gabizon2019plonk,chiesa2020marlin,bunz2020transparent} to prove the end-to-end training process in FL is challenging. This is because \first the detailed local model evaluation algorithm can be complex and even contains computations over homomorphically encrypted values (see \S~\ref{subsec:model-aggr}); and \second the model sizes are large, for instance, with millions of floating-point parameters or even more. Both of these issues would result in significantly large proof circuits, which are impractical to implement.

%% file: 3.framework.tex
\section{\sys Overview}

Architecturally, \sys is designed around four components (as shown in Figure~\ref{fig:arch}(b)). \first A data acquirer (\dacquirer) relies on \sys to collect training data for a FL training task from a utility-driven marketplace like \sys. Each training epoch is associated with a reward that the \dacquirer will pay after the data trading is closed. 
\second Data providers (\dproviders) participate in FL training by contributing their local model updates. 
\sys itself has two building blocks. \third A Quality-aware Model Evaluation Protocol that enables the \dacquirer to confidentially pre-evaluate the quality of the local models from different \dproviders. The \dacquirer can keep the detailed aggregation algorithm (\eg how to remove poisonous local models) confidential, making it difficult for the malicious \dproviders to manipulate the training process (see analysis in \S~\ref{subsubsec:aggregation_analysis}).  
\fourth Afterwards, they apply the Verifiable Transaction Protocol to achieve fair data trading. The \dacquirer first commits the \emph{aggregation weights}, obtained by the model evaluation protocol, on the trading smart contract. Upon commitment, the \dproviders can safely submit their plaintext local models offline to the \dacquirer. The \dacquirer is expected to generate a publicly verifiable proof (without disclosing its model evaluation method and the received local models) to demonstrate that it has faithfully aggregated these local models using the committed weights. Given the proof, the \dproviders can unambiguously claim the reward (proportional to their aggregation weights) deposited by the \dacquirer on the smart contract. Violations against the transaction protocol (\eg the proof verification fails) results in automatic penalties coded in the smart contract.

%% file: 4.model_evaluation.tex
\section{Quality-Aware Model Evaluations}
\label{sec:model-aggr}

\begin{figure}[!t]
    \centering  
    \subfigcapskip=-2pt 
    \subfigure[Targeted Attack]{
        \includegraphics[width=0.48\linewidth]{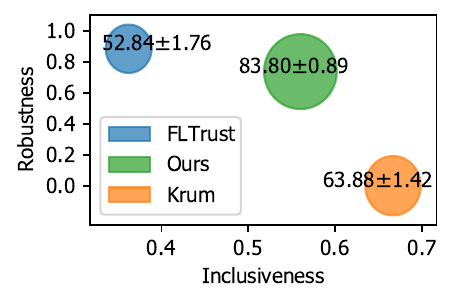}}
    \subfigure[Untargeted Attack]{
        \includegraphics[width=0.48\linewidth]{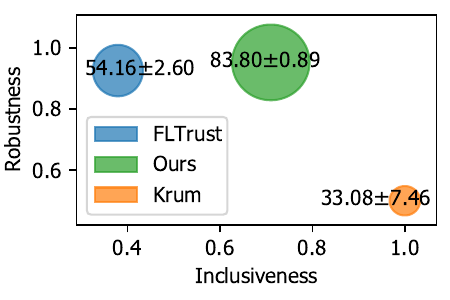}}
    \caption{The tradeoff between robustness and inclusiveness in prior robust FL approaches.}
 \label{fig:tradeoff}
\end{figure}

\subsection{Key Observations} 
\label{subsubsec:key_observation}
We first discuss the key observations about the tradeoff between the inclusiveness and robustness in the prior client-driven and server-driven secure FL aggregation protocols, which motivates our model aggregation design.
We consider a common data trading scenario where the \dacquirer has an unevenly distributed root dataset prior to trading, and the data qualities for different \dproviders vary and some \dproviders are malicious. Specifically, using the TREC dataset~\cite{li2002learning} as an example, suppose that \first the root dataset of the \dacquirer is dominated by half of the class labels; \second the \dproviders are heterogeneous, where 30\% of them have high-quality data (evenly distributed across all types of labels), 30\% of them own biased dataset, and 40\% of them are malicious; \third the malicious \dproviders may launch the backdoor attack \cite{bagdasaryan2020backdoor} (a type of the targeted attack) or the sign-randomizing attack (a type of untargeted attack). 
We evaluate two representative prior art using this setting: a server-driven design FLTrust~\cite{cao2021fltrust} and a client-driven design Krum~\cite{blanchard2017machine}.

We report three metrics in Figure \ref{fig:tradeoff}. 
Robustness represents the ability to exclude poisoned local models, quantified by the percentage of malicious \dproviders whose local models are not selected for aggregation. 
Inclusiveness represents the ability to identify benign \dproviders, quantified by the percentage of benign \dproviders whose local models are selected for aggregation. Accuracy represents the final model performance on the testset. 
We observe a clear tradeoff between inclusiveness and robustness in prior art, where the server-driven approach has higher robustness while only selecting \dproviders similar to the (biased) root dataset (sacrificing inclusiveness), and the client-driven design behaves the opposite.  Instead, our design strikes a good balance between robustness and inclusiveness, thus yielding significant accuracy gain over prior art. 
In \S~\ref{subsubsec:tradeoff_analysis}, we further investigate this tradeoff using a series of different parameters.

\subsection{Model Aggregation Protocol}
\label{subsec:model-aggr}

\input{algorithm}


The architecture of our local model evaluation protocol is presented in Algorithm \ref{alg:aggregation}. The \dacquirer first prepares a baseline model using its own root dataset (could be biased). This model will be used as a reference for scoring other local models submitted by \dproviders in each training epoch. Afterwards, the \dacquirer clusters the \dproviders according to their scores and removes the outliers (\ie low-quality local models) for this epoch. Finally, the \dacquirer and selected \dproviders finalize the data trading using our verifiable transaction protocol detailed in \S~\ref{sec:verifiable_transaction}, which guarantees that the \dacquirer distributes rewards to the \dproviders according to their model quality. 
Before starting the next training epoch, the \dacquirer \emph{dynamically adjusts} the baseline to incorporate the high-quality data collected in the prior epoch, which is the key to address the possibly biased root dataset. Throughout the training process, we apply Homomorphic Encryption (HE) to ensure that the \dacquirer cannot obtain plaintext local models before committing to purchase them.

\subsubsection{Hierarchical Clustering for Outlier Removal}
\begin{figure}[!t]
\centering
\includegraphics[width=\columnwidth]{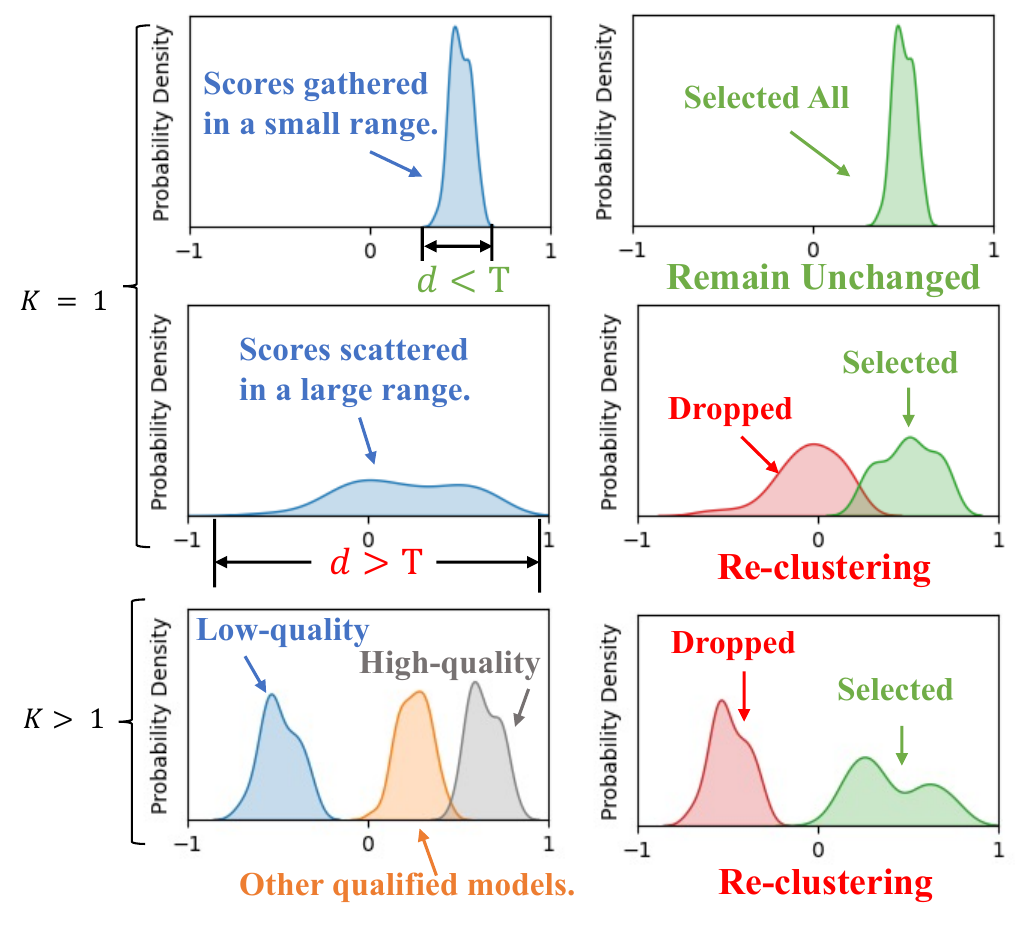}
\caption{Three different cases of distribution of scores.}
\label{fig:cluster}
\end{figure}

We score the local models submitted by \dproviders using cosine similarities (similar to FLTrust~\cite{cao2021fltrust}). 
Suppose that $W_g^t$ is the global model at round $t$, 
$W_g^{t'}$ is the baseline model (in the first epoch, it trained from $W_g^t$ by the \dacquirer with its root dataset $D_0$),
and $u_g^t = \textsf{Flatten}(W_g^{t'}-W_g^t)$ is therefore the self-update computed by the \dacquirer. 
Suppose that $W_i^t$ is the model obtained by the $i$-th \dprovider after it trains $W_g^t$  on its local dataset $D_i$, and $u_i^t = \textsf{Flatten}(W_i^t-W_g^t)$ is the update computed by the $i$-th \dprovider. 
Then, the score of $u_i^t$ is calculated as follows. 
\begin{equation}
s_i^t = \text{Cosine}(u_g^t,u_i^t) = \frac{u_g^t \cdot u_i^t}{\vert \vert u_g^t \vert \vert  \cdot \vert \vert u_i^t \vert \vert}
\label{equ:cosine}
\end{equation}
The \dacquirer selects the desired updates according to their scores. Unlike the FLTrust~\cite{cao2021fltrust} that simply clips the scores via ReLU, our design avoids simply referencing the \dacquirer's root dataset by analyzing the cluster distribution of all scores. Specifically, we propose a hierarchical clustering algorithm to select the desired updates. We first apply the Gap-Statistics algorithm~\cite{tibshirani2001estimating} to determine the optimal number of clusters $\hat{g}$ \footnote{The elbow coefficient \cite{thorndike1953belongs} and silhouette \cite{rousseeuw1987silhouettes} are other possible methods to calculate $\hat{g}$. However, the elbow coefficient algorithm requires manual judgment to determine the position of the elbow, and the silhouette algorithm can only be used with two or more clusters.}. Afterwards, we obtain our first-layer clustering by applying the K-Means algorithm \cite{lloyd1982least} with $\hat{g}$. This may produce three types of distributions, as shown in Figure~\ref{fig:cluster}.



\begin{itemize}[leftmargin=*]
    \item Single-cluster gathered distribution: the model scores are concentrated, and the range of scores is less than a predefined threshold $T$. This often indicates models submitted by all \dproviders have comparable quality. 
    In this case, we include all updates to the high-quality model set $\mathcal{P}_1$ (line~\ref{line:single-gathered} of Algorithm~\ref{alg:aggregation}).
    \item Single-cluster scattered distribution: the model scores are scattered over a large range. This could be a sign of attack, where the malicious \dproviders intentionally submit arbitrary model updates. As a result, we perform the second layer of clustering (via K-Means clustering with $\hat{g}{=}2$) to divide the scores into a high-quality cluster and a low-quality cluster. The updates in the high-quality cluster are selected for aggregation in this round  (line~\ref{line:high-qulity} in Algorithm~\ref{alg:aggregation}) . 
    \item Multi-cluster distribution: the update scores form multiple clusters. This could be caused by highly-heterogeneous \dproviders where some of them possess good dataset, some of them possess biased dataset and some of them are malicious. The algorithm performs the second clustering by separating these first-stage clusters into two categories. The low-quality category is discarded. Within the high-quality category, the updates in the highest-score cluster are added to the set $\mathcal{P}_1$. The local models in the remaining clusters of the high-quality category considered to be qualified (line~\ref{line:qualified_model} of Algorithm~\ref{alg:aggregation}), but weighted based on their distances to the centroid of the highest-score cluster. 
\end{itemize}

Eventually, the \dproviders in $\mathcal{P}_1$ and a small subset of \dproviders (\eg 5\%-10\%) randomly selected from $\mathcal{P}_2$ are selected for aggregation. The \dacquirer first commits to purchase local models from these \dproviders. Afterwards, it is safe for the selected \dproviders to hand over the plaintext local models to the \dacquirer (see the detailed transaction protocol in \S~\ref{sec:verifiable_transaction}). 

\subsubsection{Dynamic Baseline Adjustment}
\label{subsec:dynamic_selection}

To avoid overfitting to the \dacquirer's root dataset, \sys enables the \dacquirer to dynamically adjust the baseline for outlier removal. Specifically, for each local model $m \in \mathcal{M}^t$, the \dacquirer evaluates $m$ on its root dataset and computes the Kappa coefficient~\cite{cohen1960coefficient}. The \dacquirer then selects the \dproviders with high Kappa coefficients as \emph{preferred \dproviders}. 
For simplicity, Algorithm~\ref{alg:aggregation} only selects the \dprovider with the highest-Kappa-coefficient as the single preferred \dprovider (line~\ref{line:preferred_dp}). 
In the next epoch, the \dacquirer trades the local models in advance with these preferred \dproviders and aggregates them as the new baseline. The \dacquirer should not disclose these preferred \dproviders until they have committed their local models.

\subsection{The Integrated Training Process}
\label{subsec:training_process}

We have described the local model evaluation protocol in \sys. In a utility-driven data marketplace, it is critical to ensure that the \dacquirer cannot obtain the plaintext model updates before committing to purchase them. 
Towards this end, we apply the CKKS Homomorphic Encryption \cite{cheon2017homomorphic} to allow \dacquirer to privately assess the local models submitted by the \dproviders.


Supposed that in the $t$-th epoch, the \dacquirer obtains the baseline update $u_g^t$. Instead of directly sharing $u_g^t$ with the \dproviders, the \dacquirer homomorphically encrypts it by the public key $k$ as $c_g^t = \textsf{Enc}(k,\frac{u_g^t}{\vert \vert u_g^t \vert \vert})$. Once a \dprovider receives $c_g^t$, it multiply its local update $u_i^t$ with $c_g^t$ as $c_i^t = \frac{u_i^t}{\vert \vert u_i^t \vert \vert} \cdot c_g^t$, and returns the result back to the \dacquirer. Eventually, the \dacquirer receives the encrypted cosine $c_i^t$, and decrypts it to obtain the score for the update $u_i^t$. Afterwards, the \dacquirer can perform local model evaluations as described in \S~\ref{subsec:model-aggr}.

A critical step in adopting CKKS is to safeguard against the \dproviders from using different models in model evaluation and subsequent model transactions, \ie preventing the \dproviders from intentionally submitting different model updates after being selected by the \dacquirer. 
To this end, we require the \dproviders to commit their model updates before model evaluations. These committed updates are then used to ensure the correctness of subsequent model transactions, as we will further discuss below.

%% file: algorithm.tex
\begin{algorithm}[t]
    \small
    \DontPrintSemicolon
    
    \textbf{\textsf{Inputs: }} 
    The scores of local models in the  $t$-th training epoch $\mathcal{S}^t = \{s_1^t, s_2^t, \ldots, s_n^t\}$; the \dprovider selected as the baseline in the $t$-th epoch $p^t$; a control flag $\alpha$ for baseline adjustment; the ratio of randomly selected baseline candidates $\beta$; the threshold $T$ used in hierarchical clustering; the root dataset $D_0$; the maximum number of clusters $G$. \\
    \textbf{\textsf{Outputs: }} The aggregation weights obtained for the $t$-th epoch and the \dprovider selected as the baseline for the $(t+1)$-th epoch $p^{t+1}$.
    
    \hrulefill
    
    \func \textsf{\textbf{Main}} $(\mathcal{S}^t,p^t,\alpha,\beta,T,D_0,G)$ \bco \\
    \mycomment{ // Set $\mathcal{P}^t$ stores the \dproviders selected for aggregation; Set $\mathcal{K}^t$ stores their weights.}\\ 
    $\mathcal{P}^t, \mathcal{K}^t \gets$ \textsf{OutlierRemoval} $(\mathcal{S}^t,p^t,\beta,T,G)$ \\ 
    \mycomment{ // $\mathcal{M}^t$ are the plaintext models that the \dacquirer commits to purchase.} \\
    $\mathcal{M}^t \gets$ \textsf{ModelTrading}$(\mathcal{P}^t)$\\
    
    \lIf{$\alpha$ = \textnormal{\textbf{true}}}{$p^{t+1} \gets$ \textsf{BaselineAdjustment}$(\mathcal{M}^t, D_0)$}
    \lElse{$p^{t+1} \gets 0$}
    
    \hrulefill
    
    \func \textsf{\textbf{OutlierRemoval}}$(\mathcal{S}^t,p^t, \beta,T,G)$ \bco \\
    $\mathcal{U} \gets \{1,2,\ldots,n\}$, $\mathcal{P}_1 \gets \emptyset$, $\mathcal{P}_2 \gets \emptyset$, $\mathcal{K} \gets \{1.0,\ldots,1.0\}$ \\
    \mycomment{// Determine the number of clusters $\hat{g}$ by Gap statistics.}\\
    \For{$g \gets 1,2,\ldots, G$}{
        $\hat{g} \gets \text{the minimum g such that } \textsf{Gap}(g) - \textsf{Gap}(g+1) + \sigma_{g+1} \geq 0$
    }
    $d \gets $ \textsf{Max}$(\mathcal{S}^t)-$\textsf{Min}$(\mathcal{S}^t)$\\
    \lIf{$\hat{g} = 1$ \textnormal{\textbf{and}} $d > T$}{$\hat{g} \gets 2$}
    \lElse{$\mathcal{P}_1 \gets \mathcal{U}$ \mycomment{ // Single-cluster gathered distribution.}}\label{line:single-gathered}
    \mycomment{ // \textsf{K-Means} returns the clusters and centroids of the scores.}\\
    $\mathcal{N}_1,\mathcal{C}_1 \gets $ \textsf{K-Means}$(\mathcal{S}^t,\hat{g})$\\ 
    $C_{best} = \textsf{Max}(\mathcal{C}_1)$ \mycomment{ // Centroid of the highest-score cluster.}\\
    \lIf{$\hat{g} > 2$}{$\mathcal{N}_2, \mathcal{C}_2 \gets$ \textsf{K-Means}$(\mathcal{S}^t,2)$ \mycomment{// Re-clustering.}} 
    \lElse{$\mathcal{N}_2 \gets \mathcal{N}_1$}
    \For{$i \gets 1,2,\ldots,n$}{
        \lIf{$\hat{g} = 1$}{\textbf{\textnormal{break}}}
        \If{$i = p^t$ \textnormal{\textbf{or}} $\mathcal{N}_1[i] = 0$ \textnormal{\textbf{or}} $\mathcal{N}_2[i] = 0$}{
            $\mathcal{K}[i] \gets 0.0$ \mycomment{ // Low-quality model.}\label{line:low-qulity}\\}
        \ElseIf{$\mathcal{N}_1[i] = \hat{g} - 1$ \textnormal{\textbf{and}} ($\mathcal{N}_2[i] \neq 0$ )}{
            $\mathcal{K}[i] \gets 1.0$, $\mathcal{P}_1$.add($i$) \mycomment{ // High-quality model.}\label{line:high-qulity}\\
        }}
    \Else{
            $\mathcal{K}[i] \gets 1.0 - \frac{\textsf{Abs}(\mathcal{S}[i] - C_{best})}{\textsf{Max}(\textsf{Abs}([s_i^t -C_{best} \textnormal{\textbf{ for }} s_i^t \textnormal{\textbf{ in }} \mathcal{S}]))}$ \label{line:qualified_model}\\
            $\mathcal{P}_2$.add($i$)\mycomment{ // Qualified but weighted model.}\\
    }
    
    \If{$ \mathcal{P}_2 = \emptyset$ \textnormal{\textbf{and}} \textsf{\textnormal{Len}}$(\mathcal{P}_1) < 0.5 \times n$}{$\mathcal{P}_2 \gets$ \textsf{RandomSample}$(\mathcal{U}-\mathcal{P}_1,\beta)$} \label{line:random_sample}

    \return $\mathcal{P}_1 \bigcup \mathcal{P}_2$, $\frac{\mathcal{K}}{\textsf{Sum}(\mathcal{K})}$
    
    \hrulefill 
    
    \func \textsf{\textbf{BaselineAdjustment}}$(\mathcal{M}^t,D_0)$ \bco \\
    $kp_{max}$ = \textsf{-inf}, $p^{t+1} = 0$\\
    \For{$i, m$ \textnormal{\textbf{in}} \textnormal{\textbf{Enumerate}}$(\mathcal{M}^t)$}{
        $kp \gets \textsf{Kappa}(m, D_0)$ \\
        \lIf{$kp > kp_{max}$}{$kp_{max} \gets kp, p^{t+1} \gets i$}\label{line:preferred_dp}
    }
    \return $p^{t+1}$
    \caption{Quality-Aware Model Aggregation Protocol}
    \label{alg:aggregation}
\end{algorithm}

%% file: 5.verifiable_transaction.tex
\section{Verifiable Transaction Protocol}
\label{sec:verifiable_transaction}

Our verifiable transaction protocol has two phases: \first a zero-knowledge proving system that allows the \dacquirer to prove that it has faithfully aggregated the global model based on claimed weights, without disclosing the local models submitted by the \dproviders; and 
\second a payment protocol based on smart contract to allow the \dacquirer and \dproviders to exchange rewards and plaintext local models. 

\subsection{Proving Scheme for Model Aggregation}
\label{subsec:verifiable_agg}


\subsubsection{Overview}  
The \dacquirer should prove that it faithfully aggregates the global model. Although there are many zero-knowledge proof (ZKP) constructions~\cite{libsnark2020, arkworksrs2023, eberhardt2018zokrates}, it is challenging to simply adopt these designs to achieve verifiable aggregation in \sys. Specifically, the model evaluation algorithm (Algorithm~\ref{alg:aggregation}) used by \sys is complex, especially considering the homomorphic computations involved. This complexity makes it difficult to generate and implement the arithmetic circuit to represent the algorithm. To address this challenge, \sys does not prove the end-to-end training process. Instead, it only proves the local model summation computation, which aggregates the local models using the aggregation weights returned by the model evaluation algorithm. This design drastically reduces the proving complexity without affecting the fairness of billing, because reward allocations are completely driven by the aggregation weights. 
In addition, we also design verifiable sampling method such that the \dacquirer only needs to prove a fix number of scalars/parameters regardless of the model size (\ie the number of model parameters).

\parab{Setup.}
Denote the local model summation as $\mathcal{A}$, which represents the following calculation $W_g^t = W_g^{t-1} + K^tU^t$, where $W_g^{t-1}$ ($W_g^t$) is the global model in the previous (current) epoch. For each $k_i^t \in \mathcal{K}^t$, $K^t = [k_1^t,k_2^t,\dots,k_n^t]$ are the aggregation weights claimed by the \dacquirer, and $U^t = [u_1^t,u_2^t,\dots,u_n^t]$ are the local models submitted by the \dproviders. 
The public input of our zero-knowledge proving scheme is $\mathcal{X}^t = \{W_g^{t},W_g^{t-1},K^t\}$, and the private witness is $\mathcal{W}^t = \{U^{t} \}$. Concretely, our proving scheme has the following algorithms.

\begin{itemize}[leftmargin=*]
    \item $\mathbb{C} \gets$ \textsf{Compile($\mathcal{A}$)}: In the compiling step, the prover (\ie the \dacquirer) quantizes the floating-point public input $\mathcal{X}^t$ and private witness $\mathcal{W}^t$ to $\mathbb{X}^t = \{\mathbb{W}_g^{t},\mathbb{W}_g^{t-1},\mathbb{K}^t\}$ and $\mathbb{W}^t = \{\mathbb{U}^{t}\}$ in finite field, respectively. In addition, it quantizes the aggregation algorithm $\mathcal{A}$ and compiles it to a circuit $\mathbb{C}$.
    \item $(pk, vk) \gets $\textsf{Setup($1^{\lambda},\mathbb{C}$)}: Given a security parameter $\lambda$ and the circuit $\mathbb{C}$, a trusted third party randomly generates a proving key $pk$ and a verification key $vk$. The proving key $pk$ is given to the \dacquirer and the verification key $vk$ is given to the \dproviders. We consider a proving scheme that requires trusted setup in this paper, and leave exploration of trust-free schemes in future work.
    
    \item $(cm^t, \mathbb{W}_g^t, \pi^t) \gets$ \textsf{Prove($\mathbb{X}^t, \mathbb{W}^t, r^t, pk, \mathbb{C}$)}: Given a random opening $r^t$, the prover first commits the private witness $\mathbb{W}^t$ as $cm^t = \textsf{Commit}(\mathbb{U}^t,r^t)$. Then it calculates the quantized global model $\mathbb{W}_g^t$ and generates a proof $\pi^t$. Afterwards, the prover publishes $cm^t$, $\pi^t$, and $\mathbb{W}_g^t$ to the \dproviders.
    
    \item $\{1,0\} \gets $ \textsf{Verify($\mathbb{X}^t, vk, \pi^t, cm^t$)}: The public verifiers (\eg \dproviders) can verify the computation in $\mathbb{C}$ using the verification key $vk$, public input $\mathbb{X}^t$, the commitment $cm^t$, and the proof $\pi^t$. If the \dacquirer faithfully aggregates the global model, the verifier will accept the proof; otherwise, the verifier will reject it.
\end{itemize}



\subsubsection{Circuit Design}

\begin{figure}[t]
\centering
\includegraphics[width=\columnwidth]{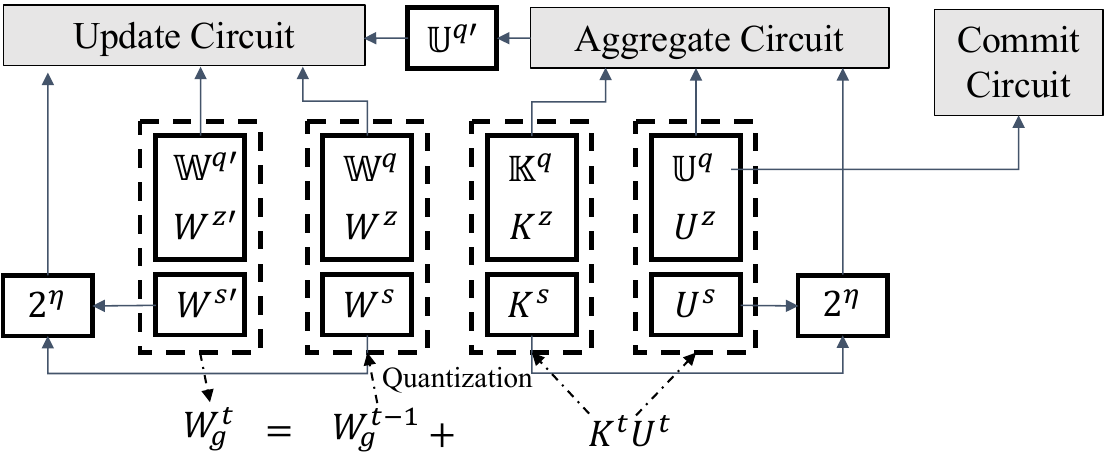}
\caption{The circuit design for the proving scheme in \sys.}
\label{fig:circuit}
\end{figure}


\parab{Quantization}
The circuit $\mathbb{C}$ is designed based on the quantized version of algorithm $\mathcal{A}$. Quantization maps a floating point value $x \in [a,b]$ to an unsigned integer $x_q \in [a^q,b^q]$ and de-quantization is the reversed process. As defined in the partial quantization~\cite{jacob2018quantization}, the quantization and de-quantization are represented as $q = \lfloor\frac{x}{s}\rfloor + z$ and $x = s(q-z)$, respectively, where $s$ is a floating-point scaler, $q$ is the quantized integer for $x$, and $z$ is the zero point (\ie the value of a floating-point zero when mapped to the integer field). 

We observe in our experiments that the above quantization design may result in overflow. Specifically, the subtractions on unsigned integers may result in overflow due to accuracy loss in quantization. Consequently, the de-quantization produces a very inaccurate dequantized model for the next training epoch.   
To avoid overflow, we extend the range of floating-point numbers by a small $\epsilon$, \ie $x \in [a - \epsilon, b + \epsilon]$. Afterwards, we derive $s$ and $z$, by solving the following linear Equation (\ref{equ:s_z_linear}).


\begin{equation}
\label{equ:s_z_linear}
    a - \epsilon  = s (a^q + z) ; \quad 
    b + \epsilon  = s (b^q + z)    
\end{equation}

\parab{Commitment Circuit.} 
The first part of $\mathbb{C}$ is to commit the private parameters $\mathbb{U}^t$ with the opening random $r^t$ such that $\mathbb{U}^t$ are not disclosed to the public verifiers, \ie $cm^t = \textsf{Commit}(\mathbb{U}^t,r^t)$. 
POSEIDON~\cite{lorenzo2021poseidon} is an optimized commitment algorithm. Yet, simply applying POSEIDON to commit all model parameters would require in a large number of constraints. In \S~\ref{subsubsec:sampling}, we design a verifiable sampling mechanism to avoid committing and verifying all parameters.  


\parab{Aggregation Circuit Design.} 
The second part of $\mathbb{C}$ is the aggregation circuit that computes 
$U^{t'}= K^tU^t$ in the quantized form, where $K^t \in \mathbb{R}^{1 \times n}$, $U^t  \in \mathbb{R}^{n \times m}$, $U^{t'} \in \mathbb{R}^{1 \times m}$, $n$ is the number of \dproviders, and $m$ is the number of parameters in the model. To be ZKP-friendly, we minimize the use of negative numbers and division in the calculation, while ensuring that all operations are performed in the field $\mathbb{F}_q$. 
First, in Equation (\ref{equ:aggregate1}), we apply the de-quantization equation. 
\begin{equation}
\label{equ:aggregate1}
U^{s'}(\mathbb{U}^{q'}_{i,j} - U^{z'}) = \sum_{k=1}^{n} K^s(\mathbb{K}^q_{i,k} - K^z) U^s(\mathbb{U}^q_{k,j} - U^z)
\end{equation}
where $\mathbb{K}^q$, $\mathbb{U}^q$ and $\mathbb{U}^{q'}$ are the quantization matrix of $K^t$, $U^t$, and $U^{t'}$, respectively;  $K^s$, $U^s$ and $U^{s'}$ are the scaler of $K^t$, $U^t$, and $U^{t'}$, respectively; $K^z$, $U^z$ and $U^{z'}$ are the zero points of $K^t$, $U^t$, and $U^{t'}$, respectively. In the context, $\mathbb{K}^t = \{\mathbb{K}^q,K^s,K^z\}$, etc.
We use a big integer $2^\eta$ ($\eta$ should be $22$ or even larger) to replace the floating-point scale with unsigned integers and enable the full quantization computation. Also, we rearrange the calculation order in Equation (\ref{equ:aggregate2}) to eliminate negative numbers in calculation. 
The remainder $\mathbb{R}^a$ is to ensure correctness after  division, as shown in~\cite{feng2021zen}.


\begin{equation}
\label{equ:aggregate2}
\begin{split} 
2^{\eta} \mathbb{U}^{q'}_{i,j} & =  \mathbb{R}^a_{i,j} +  2^{\eta} U^{z'} + \bigg\lfloor 2^{\eta} \frac{K^s U^s}{U^{s'}} \bigg( M_1 + M_4 - M_2 - M_3\bigg) \bigg\rfloor \\
\text{ s.t. } \xspace M_1 & = \sum_{k=1}^{n} \mathbb{K}^q_{i,k} \mathbb{U}^q_{k,j}, M_2 = U^z \sum_{k=1}^{n} \mathbb{K}^q_{i,k},\\
M_3 & = K^z \sum_{k=1}^{n} \mathbb{U}^q_{k,j}, M_4 = n K^z U^z\\
\end{split}
\end{equation}

\parab{Update Circuit Design.} 
The third part of $\mathbb{C}$ is the update circuit. We use Equation (\ref{equ:update1}) to  present the de-quantized update equation $W_g^t = W_g^{t-1} + U^{t'}$, with $W_g^t \in \mathbb{R}^{1 \times m}$, $W_g^{t-1} \in \mathbb{R}^{1 \times m}$, $U^{t'} \in \mathbb{R}^{1 \times m}$.

\begin{equation}
\label{equ:update1}
W^{s'}(\mathbb{W}^{q'}_{i,j} - W^{z'}) = W^s(\mathbb{W}^q_{i,j} - W^z) + U^{s'}(\mathbb{U}^{q'}_{i,j} - U^{z'}),
\end{equation}
where $\mathbb{W}^{q'}$, $\mathbb{W}^q$ and $\mathbb{U}^{q'}$ are the quantization matrices of $W_g^t$, $W_g^{t-1}$ and $U^{t'}$, respectively; $W^{s'}$, $W^s$ and $U^{s'}$ are the scaler of $W_g^t$, $W_g^{t-1}$ and $U^{t'}$, respectively. $W^{z'}$, $W^z$ and $U^{z'}$ are the zero points of $W_g^t$, $W_g^{t-1}$ and $U^{t'}$, respectively.

Similarly, we rearrange the above equation to Equation (\ref{equ:update2}) to eliminate negative numbers. And the remainder $\mathbb{R}^u$ to ensure correctness after division.

\begin{equation}
\label{equ:update2}
\begin{split} 
2^{\eta} \mathbb{W}^{q'}_{i,j} & =\mathbb{R}^u_{i,j} + 2^{\eta} W^{z'} + \bigg\lfloor 2^{\eta}  \bigg(N_1+N_3-N_2-N_4\bigg)\bigg\rfloor\\
\text{ s.t. }N_1 & = \frac{W^s}{W^{s'}}\mathbb{W}^q_{i,j}, N_2 = \frac{W^s}{W^{s'}}W^z, \\
N_3 & = \frac{U^{s'}}{W^{s'}}\mathbb{U}^{q'}_{i,j}, N_4 = \frac{U^{s'}}{W^{s'}}U^{z'}
\end{split}
\end{equation}

In summary, the complete circuit $\mathbb{C}$ is plotted in Figure~\ref{fig:circuit}. 

\subsubsection{Verifiable Sampling} 
\label{subsubsec:sampling}

Given the concatenated local models $U^t \in \mathbb{R}^{n \times m}$ 
(where $n$ is the number of \dproviders, and $m$ is the number of parameters in the model), the number of constraints required in the commitment circuit, the aggregation circuit, and the update circuit  is $\mathcal{O}(H \cdot n \cdot m)$,  $\mathcal{O}(n \cdot m)$ and $\mathcal{O}(m)$, respectively, where $H$ represents the required constraints in the hash function used in commitment circuit. Considering that $n \ll m$ and $H$ is fixed once the commitment hash function is selected, we explore to reduce the number of parameters required for proof generation. Specifically, we randomly select $c$ out of $m$ parameters as the verification objects. Suppose that the sampling is provable random (\ie not controlled by the \dacquirer), as long as the \dacquirer can provide the correct proof for the sampled parameters, then with high probability, the \dacquirer has calculated all parameters correctly. Thus, the proof complexity becomes independent on $m$. 

Conceptually, the provable random sampling is similar to randomness beacon~\cite{rabin1983transaction}. Both verifiable random function (VRF)~\cite{micali1999verifiable,algorand_vrf} and verifiable delay function (VDF)~\cite{boneh2018verifiable} can be used as a primitive to construct the verifiable random sampling. We sketch a construction below. In each training epoch, each \dprovider publishes a  cryptographic nonce to a public bulletin board (\eg a public blockchain). 
The \dacquirer is required to use $\mathcal{H}(s_1, s_2, ..., s_n)$ as the seed $s^t_{\textsf{vdf}}$ to a \textsf{VDF} to select the parameter indices $R^t_{\textsf{vdf}} = \{r_1^t,r_2^t,\dots,r_c^t\}$ (\eg using the output of the \textsf{VDF} as the random seed for a pre-agreed pseudorandom number generator). \textsf{VDF} is necessary to prevent the \dprovider that lastly publishes its nonce from introducing bias by strategically selecting its nonce. 

After random sampling, the public input and private witness for the proving scheme should be also adjusted accordingly as $\mathbb{W}_g^{t,c}=\{\mathbb{W}_{g,r_1^t}^t, \mathbb{W}_{g,r_2^t}^t, \dots, \mathbb{W}_{g,r_c^t}^t\}$, $\mathbb{W}_{g}^{t-1,c} = \{\mathbb{W}_{g,r_1^t}^{t-1}, \mathbb{W}_{g,r_2^t}^{t-1}, \dots, \mathbb{W}_{g,r_c^t}^{t-1}\}$, $\mathbb{U}^{t,c} = \{u_{r_1^t}^t, u_{r_2^t}^t, \dots, u_{r_c^t}^t \}$, where $ \mathbb{W}_g^{t,c} \in \mathbb{R}^{1 \times c}$, $\mathbb{W}_g^{t-1,c} \in \mathbb{R}^{1 \times c}$, $\mathbb{U}^{t,c} \in \mathbb{R}^{n \times c}$.
As a prerequisite for using the sample-based verification, the \dacquirer shall publish the model $\mathbb{W}_g^t$ before sampling (since only part of the $\mathbb{W}_g^t$ is used as the public input). 


\subsubsection{Integrated Verification Protocol}

Taken all parts together, our verifiable aggregation protocol proceeds as follows. 
The \dacquirer first quantizes $W_g^{t-1}$, $U^t$ and $K^t$ to the quantized format, and performs the aggregation calculation as Equation (\ref{equ:aggregate2}) and Equation (\ref{equ:update2}). In addition, the \dacquirer applies the de-quantization equation to calculate the floating-point global model $W_g^t$, and commits $K^t, W_g^t$ and $\mathbb{W}_g^t$ to \dproviders. 
Afterwards, the \dacquirer obtains the randomly selected parameters from the \textsf{VDF}, and generates a zero-knowledge proof $\pi^t$ with public input as $\mathbb{X}^{t,c} = \{ \mathbb{W}_g^{t,c}, \mathbb{W}_{g}^{t-1,c}, \mathbb{K}^{t} \}$ and private witness as $\mathbb{W}^{t,c} = \{\mathbb{U}^{t,c}\}$. The proof $\pi^t$ is then submitted to a smart contract so that the \dproviders can verify its correctness and claim corresponding rewards (see \S~\ref{subsec:payment}). 

\input{smart_contract}

%% file: smart_contract.tex
\subsection{The Trading Smart Contract}
\label{subsec:payment}

\sys designs a trading smart contract to enable the \dacquirer and \dproviders to exchange plaintext local models and rewards. 
Due to space constraint, we provide the high-level description of our smart contract in Algorithm~\ref{alg:contract}. 
The more detailed realization of our smart contract that is close to the real-world implementation is deferred to \S~\ref{subsec:pseudocode}.
The trading smart contract is divided into two high-level phases. 


\begin{algorithm}[t]
\caption{The Trading Smart Contract}
\label{alg:contract} 
    \small
    \preparephase() \bco \\
    \quad commit $K^t$, $\textsf{addrs}$ {\color{comment} \# $\textsf{addrs}$ identify selected \dproviders in current epoch}\\
    \quad $v_{\textsf{DPs}},v_{\textsf{DA}}$ = \deposit(\msg.value) {\color{comment} \# \dacquirer deposits (reward, penalty)}\\
    \quad $R_{\textsf{DPs}} := $ \textsf{Allocate}($v_{\textsf{DPs}}$, $K^t$) {\color{comment} \#  allocate \dproviders reward based on $K^t$} \\
    \quad $U^t :=$ Submission($\textsf{addrs}$) {\color{comment} \# \dproviders submit local models off-chain}\\
    \quad $\mathbb{W}_g^t :=$ Aggregate($\mathbb{W}_g^{t-1}, \mathbb{K}^t, \mathbb{U}^t$) {\color{comment} \# \dacquirer generates proof off-chain}\\
    \quad commit $\mathbb{W}_g^t,\mathbb{W}_g^{t-1}, \mathbb{K}^t$  {\color{comment} \# \dacquirer commits public inputs}\\
    
    \verifyphase() \bco \\
    \quad \dacquirer performs verifiable sampling offline and publishes  $s^t_{\textsf{vdf}},\pi^t_{\textsf{vdf}}$ \\
    \quad \dacquirer adjusts $\mathbb{W}_g^{t,c}, \mathbb{W}_{g}^{t-1,c}$ and $\mathbb{U}^{t,c}$ based on $R^t_{\textsf{vdf}}$ \\
    \quad \dacquirer generates $\pi^t_{\textsf{agg}} := $Prove($\mathbb{X}^{t,c},\mathbb{W}^{t,c},pk,\mathbb{C}$) offline\\
    \quad \dacquirer publishes the proof $\pi^t_{\textsf{agg}}$ on-chain \\ 
    \quad \dproviders invokes verification $v^t := $ \verify($vk$, $\mathbb{X}^{t,c}$, $\pi^t_{\textsf{agg}}$) \\
    \quad \textbf{if} $v^t = $ \fal \bco \\
    \quad \quad distribute both the security deposit $v_{\textsf{DA}}$ the award $v_{\textsf{DPs}}$ to \dproviders\\ 
    \quad \textbf{else} \bco distribute $v_{\textsf{DPs}}$ to \dproviders and return $v_{\textsf{DA}}$ to \dacquirer \\
    
    \textbf{Function} \verify($vk, \mathbb{X}^{t,c}, \pi_\textsf{agg}^t$) \bco \\   
    \quad $s := \Sigma^{\len(\mathbb{X}^{t,c})-1}_{i=0}$ \smul(vk.$\gamma_{abc}$[i + 1],$ \mathbb{X}^{t,c}$[i])\\
    \quad $s := $\add($s$, vk.$\gamma_{abc}$[0])\\
    \quad $p_1$ := $\pi_{agg}^t$.a, \negate($s$), \negate($\pi_\textsf{agg}^t$.c), \negate(vk.$\alpha$)\\
    \quad $p_2$ := $\pi_\textsf{agg}^t$.b, vk.$\gamma$, vk.$\delta$, vk.$\beta$\\
    \quad \return \pair($p_1$,$p_2$)\\

\end{algorithm}

\parab{Prepare Phase.} 
The \preparephase~performs necessary setup for reward distribution. 
First, the \dacquirer commits the aggregation weights $K^t$ and the corresponding \dproviders (identified by their public keys or addresses on blockchain) in the smart contract. Meanwhile, the \dacquirer deposits the reward $v_{\dproviders}$ for the \dproviders proportional to their weights in $K^t$.
Additionally, the \dacquirer also deposits $v_{\dacquirer}$ as the penalty if it cannot later provide a correct proof. Afterwards, the \dproviders can safely submit their plaintext local models off-chain to the \dacquirer, based on which the \dacquirer generates the verifiable proof as described in \S~\ref{subsec:verifiable_agg}. After proof generation, the \dacquirer commits the public inputs. 


\parab{Verify Phase.} 
The second phase focuses on verifying the integrity of model aggregation. The \dacquirer performs verifiable random sampling and provides the proper proof (\ie $\pi^t_{\textsf{vdf}}$) for randomness.  
Afterwards, the \dacquirer adjusts the public and private inputs according to {the random seed $s_{\textsf{vdf}}^t$, based on which it generates the final proof for model aggregation $\pi^t_{\textsf{agg}}$. The proof is uploaded to the trading smart contract such that any \dprovider can verify its correctness by invoking the on-chain \verify~function. 
The \dacquirer will lose its security deposit if the verification fails. 


\parab{On-Chain Verification Procedure.} 
The \verify~function is responsible for checking the correctness of $\pi^t_{\textsf{agg}}$. It takes input as the committed verification key $vk$ and the quantized public input $\mathbb{X}^{t,c}$, and the proof $\pi^t_{\textsf{agg}}$. 
The underlying verification is based on the \textsf{Groth16} protocol~\cite{groth2016size} which checks four pairings. The cryptography-related computations (such as $\add$ and $\smul$) are implemented via the precompiled smart contracts to reduce gas cost. 

%% file: 6.evaluation.tex
\section{Evaluation}
\label{sec:evaluation}

\subsection{Experimental Setup}

Our experiments are conducted on two Linux servers with Intel(R) Xeon(R) Gold 6348 CPU and NVIDIA RTX A100 GPU. We use Pytorch~\cite{paszke2019pytorch} to implement FL, apply SEAL~\cite{sealcrypto} for CKKS-based Homomorphic operations, and Ethereum~\cite{wood2014ethereum} testnet for deploying our trading smart contract. The source code is available at Github\footnote{\href{https://github.com/liqi16/martFL}{https://github.com/liqi16/martFL}}. All results are obtained based on five repetitions of experiments. 

\parab{Datasets, Models, and Baselines.} We use multiple datasets from different domains in our evaluations, including two image classification datasets, FMNIST~\cite{xiao2017/online} and CIFAR~\cite{krizhevsky2009learning}, and two text classification datasets, TREC~\cite{li2002learning} and AGNEWS~\cite{zhang2015character}. We train LeNet~\cite{lecun1989backpropagation} as global model for FMNIST~\cite{xiao2017/online}, TextCNN~\cite{chen2015convolutional} for TREC \cite{li2002learning} and AGNEWS \cite{zhang2015character}. We train a convolutional neural network (CNN) with three CNN layers and four linear layers as the global model for the CIFAR~\cite{krizhevsky2009learning} dataset. 
\iftechreport
The architecture of the CNN is shown in Table \ref{tab:cnncifar} of \S~\ref{subsec:cnn_arch}. 
\fi
\revision{We compare \sys with two server-driven methods (FLTrust~\cite{cao2021fltrust} and CFFL~\cite{lyu2020collaborative}) and five client-driven methods (FedAvg~\cite{mcmahan2017communication}, RFFL~\cite{xu2020reputation}, Krum~\cite{blanchard2017machine}, RLR~\cite{ozdayi2021defending}, and Median~\cite{yin2018byzantine}). }

\parab{Training.} We set the same number of participants for both client-driven and server-driven approaches. This ensures that all participants use the same number of samples in the training process. For client-driven methods, we set $n$ \dproviders. For server-driven methods, we set one \dacquirer and $n-1$ \dproviders. 
For image classification tasks, we set 30 participants, the optimizer is SGD, and the learning rate is $1.0\times10^{-2}$. For text classification tasks, we set 20 participants, the optimizer is Adam, and the learning rate is $5.0\times10^{-5}$. The number of samples in the \dacquirer's root dataset is 200 for FMNIST, CIFAR, and AGNEWS, and 120 for TREC, counting for roughly 0.3\%, 0.4\%, 2\%, and 1.6\% of the total data held by the \dproviders, respectively. Unless otherwise specified, we train a model until its peak accuracy on our validation dataset does not increase for 100 training epochs.

\parab{Data Splits.} We apply two sampling methods to divide the amount of data held by each \dprovider: \textsf{UNI} and \textsf{POW}. In \textsf{UNI}, each \dprovider has the same amount of samples; in \textsf{POW} method, the numbers of samples owned by different \dproviders follow a  power-law distribution. In addition, we divide the local data distribution of the \dproviders according to two methods, \textsf{IID} and \textsf{NonIID}. \textsf{IID} means that each \dprovider has all classes of samples and the samples in each class are uniformly distributed; \textsf{NonIID} means that the \dprovider has a subset of classes, and the data distributions vary for  different \dproviders.

\parab{The Adversary.} We consider two untargeted attacks, two targeted attacks, and Sybil attack  ~\cite{fung2020limitations}. The untargeted attacks include sign-randomizing attack and free-rider attack~\cite{lin2019free}. 
The sign-randomizing attack is an attack on the direction of the gradients where the adversary randomly sets the sign as $+1$ or $-1$. In the free-rider attack, we implement the delta weight attack \cite{lin2019free}, which generates gradient updates by subtracting the two global models received in the previous two epochs. 
The targeted attacks include label-flipping attacks and backdoor attacks~\cite{bagdasaryan2020backdoor}. In a label-flipping attack, the adversary swaps the labels of the two classes of data in the training process to train poisoned local models. In the Sybil attack, the adversary conjures up a number of clients and submit the same compromised model. In the Sybil attack, we use the label-flipping attack to train the malicious local models.

\parab{Evaluation Metrics.} We use Main Task Accuracy (\textbf{MTA}) and Attack Success Rate (\textbf{ASR}) as the evaluation metrics. MTA measures the classification accuracies of the trained models, while ASR measures the fraction of poisoned samples that are predicted as the target class in targeted attacks. Thus, higher MTAs indicate more effective models, and lower ASRs indicate more robust models against targeted attacks. 
We further define Data Acquisition Cost (\textbf{DAC}) as the average percentage of local models that the \dacquirer must procure in each training epoch in order to train the global model. In general, the \dacquirer seeks to obtain high-performing models (\ie with high MTAs and low ASRs) at a reasonable DAC (lower the better).

\parab{Default Hyper-Parameters.} 
For \sys, we set the threshold $T$ used in hierarchical clustering as $0.05$ and the ratio of randomly selected baseline candidates $\beta$ as $0.1$. 
For Krum~\cite{blanchard2017machine}, we set the proportion of possibly Byzantine as $20\%$. For CFFL~\cite{lyu2020collaborative}, we set the coefficient of reputation threshold as $1.0$ and $\alpha$ as $5$. For RFFL~\cite{xu2020reputation}, we set the hyper-parameter $\alpha$ as $0.95$ and threshold as $1.0$. 
\revision{For RLR~\cite{ozdayi2021defending}, we set the learning threshold $\theta$ is $0$.} 
For the backdoor attack, we implement the attack proposed in~\cite{bagdasaryan2020backdoor} where the hyper-parameter $\alpha$ is $0.95$.

\subsection{Evaluation Results}
\label{subsec:evaluation_results}
Our evaluations are centered around the following questions: 
\begin{itemize}[leftmargin=*]
    \item \textbf{Accuracy}. In \S~\ref{subsubsec:eval:bias} and \S~\ref{subsubsec:eval:unbias}, we quantitatively show that \sys achieves the best MTAs compared to other server-driven methods regardless of when the \dacquirer's root dataset is biased or not. Meanwhile, \sys reduces up to 69\% DAC when achieving comparable (if not better) MTAs with prior arts.
    \item \textbf{Robustness.} In \S~\ref{subsubsec:eval:robust}, we show that when facing with various targeted attacks, untargeted attacks, and Sybil attack, \sys can accurately identify malicious \dproviders and achieve the highest MTA and lowest ASR in most cases, compared with prior arts. 
    \item \textbf{Accuracy Loss by Quantization.} In \S~\ref{subsubsec:eval:quantization}, we show that quantization has little to no impact on the MTA of the global model.
    \item \textbf{System Overhead}. In \S~\ref{subsubsec:eval:efficiency}, we study the system overhead of \sys, including the cryptography overhead during local model evaluations, and the gas cost incurred for executing the trading smart contract.
\end{itemize}

\begin{table*}[ht]\small
\begin{tabularx}{\textwidth}{l l *{6}{>{\centering\arraybackslash}X}}
\toprule
\multirow{2}{*}{Dataset} & Biased Ratio & \multicolumn{2}{c}{20\%}                    & \multicolumn{2}{c}{30\%}                    & \multicolumn{2}{c}{40\%}                    \\ \cmidrule(lr){2-8}
                         & Metric       & MTA                                & DAC    & MTA                                & DAC    & MTA                                & DAC    \\ \hline
                         & CFFL         & 76.87 $\pm$ 6.87                   & 100.00 & 81.53 $\pm$ 0.90                   & 100.00 & 79.07 $\pm$ 3.24                   & 100.00 \\
TREC                     & FLTrust      & 67.40 $\pm$ 4.76                   & 36.52  & 72.60 $\pm$ 1.07                   & 46.47  & 71.73 $\pm$ 1.09                   & 40.15  \\
                         & Ours         & \textbf{88.80} $\pm$ \textbf{1.72} & 53.63  & \textbf{87.53} $\pm$ \textbf{1.15} & 53.88  & \textbf{87.20} $\pm$ \textbf{0.49} & 51.79  \\ \hline
                         & CFFL         & 44.09 $\pm$ 1.95                   & 100.00 & 43.39 $\pm$ 1.57                   & 100.00 & 45.58 $\pm$ 1.46                   & 100.00 \\
AGNEWS                   & FLTrust      & 44.09 $\pm$ 1.43                   & 11.52  & 45.19 $\pm$ 0.39                   & 10.35  & 43.65 $\pm$ 0.90                   & 11.89  \\
                         & Ours         & \textbf{79.71} $\pm$ \textbf{2.15} & 36.30  & \textbf{75.89} $\pm$ \textbf{1.30} & 38.99  & \textbf{78.04} $\pm$ \textbf{1.08} & 41.15  \\ \hline
                         & CFFL         & \textbf{88.37} $\pm$ \textbf{0.55} & 100.00 & 88.48 $\pm$ 0.25 & 100.00 & \textbf{88.02} $\pm$ \textbf{0.32} & 100.00 \\
FMNIST                   & FLTrust      & 87.33 $\pm$ 0.48                   & 32.64  & 87.26 $\pm$ 0.38                   & 33.57  & 87.28$\pm$ 0.39                    & 46.04  \\
                         & Ours         & 88.22 $\pm$ 0.26                   & 35.14  & \textbf{88.88} $\pm$ \textbf{0.27}                   & 30.06  & 87.71 $\pm$ 0.43                   & 39.60  \\ \hline
                         & CFFL         & 63.34 $\pm$ 0.22                   & 100.00 & 62.38 $\pm$ 0.33                   & 100.00 & 60.85 $\pm$ 0.80                   & 100.00 \\
CIFAR                    & FLTrust      & 10.00 $\pm$ 0.00                   & 7.59   & 11.42 $\pm$ 1.00                   & 10.79  & 14.25 $\pm$ 1.99                   & 1.30   \\
                         & Ours         & \textbf{64.24} $\pm$ \textbf{0.06} & 53.63  & \textbf{63.79} $\pm$ \textbf{0.28} & 53.88  & \textbf{62.60} $\pm$ \textbf{0.37} & 51.79  \\ \hline
\end{tabularx}
\caption{MTA (\%) and DAC (\%) when the \dacquirer possesses a biased root dataset.}
\label{tab:biased_client}
\end{table*}

\subsubsection{Biased Root Dataset} 
\label{subsubsec:eval:bias}

First, we evaluate the MTA of prior art when the \dacquirer possesses an unevenly distributed root dataset. Specifically, we consider that \first the root dataset of \dacquirer is dominated by half of the class labels; \second the \dproviders follow the form of \textsf{UNI} in the number of samples; \third a certain percentage of \dproviders have biased local dataset and the remaining \dproviders have evenly distributed dataset (\ie high-quality dataset with \textsf{IID} distributions across all class labels).

{We evaluate three different percentages of \dproviders (20\%, 30\%, and 40\%) possessing biased local datasets. The results are reported in Table~\ref{tab:biased_client}. In terms of main task accuracy (MTA), \sys consistently outperforms existing server-driven approaches, with particularly significant advantages over FLTrust. We observed that all three methods have very close MTAs on the FMNIST task. This may be because the FMNIST task is relatively simple and we use a fairly small model with approximately 44,000 parameters. The advantages of \sys become more pronounced on larger models (for instance, the models for both text classification tasks have ${\sim}3$ million parameters, and the model for the CIFAR task has ${\sim}1$ million parameters). 
The underlying reason for the MTA improvements in \sys is because existing server-driven approaches has poor inclusiveness when the root dataset is biased. To quantify this, we plot the Cumulative Distribution Function (CDF) of inclusiveness in Figure~\ref{fig:bias_data} for the TREC task with 20\% biased \dproviders. We consider both the inclusiveness of all the \dproviders and inclusiveness of only the high-quality \dproviders. 
Because existing server-driven methods tend to select local models with data distributions similar to the \dacquirer's root dataset, their selection of \dproviders is highly biased towards its root dataset. 
On the contrary, benefited from the dynamic baseline adjustment design, \sys can include more high-quality \dproviders, even if the root dataset is biased. 

We further report DACs for all three methods, which represents the average percentage of local models that the \dacquirer purchases in each training epoch. The DAC in CFFL is always 100\% because CFFL must obtain all local models and evaluate their accuracies before deciding whether or not to aggregate them. Therefore, the model aggregation design in CFFL is undesirable in data marketplace, where the \dacquirer prefers to only pay for high-quality local models from the \dproviders. 
On the contrary, FLTrust has low DACs in this setting because its local model selections are highly biased. As a result, FLTrust has the lowest MTAs in nearly all tasks. \sys instead strikes a good balance between MTA and DAC, allowing the \dacquirer to obtain high-performing global models with low cost.

\begin{figure}[!t]
    \centering  
    \subfigbottomskip=2pt 
    \subfigcapskip=-5pt 
    \subfigure[]{
        \label{fig:bias_data_a}
        \includegraphics[width=0.48\linewidth]{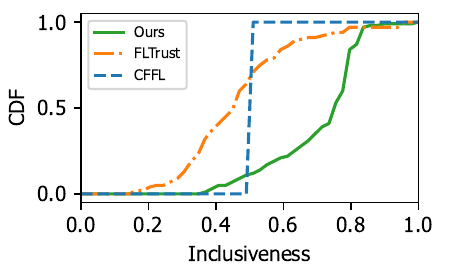}}
    \subfigure[]{
        \label{fig:bias_data_b}
        \includegraphics[width=0.48\linewidth]{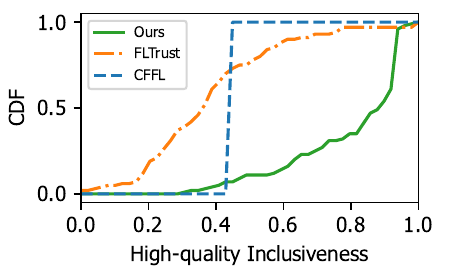}}
    \caption{The inclusiveness analysis when the \dacquirer posseses a biased root dataset.}
 \label{fig:bias_data}
\end{figure}

\begin{table*}[]\small
\begin{tabularx}{\textwidth}{
>{\raggedright\arraybackslash}X
>{\centering\arraybackslash}X
>{\centering\arraybackslash}X
>{\centering\arraybackslash}X
>{\centering\arraybackslash}X
>{\centering\arraybackslash}X
>{\centering\arraybackslash}X
>{\centering\arraybackslash}X
>{\centering\arraybackslash}X}

\toprule
Dataset & \multicolumn{2}{c}{TREC}                    & \multicolumn{2}{c}{AGNEWS}                  & \multicolumn{2}{c}{FMNIST}                  & \multicolumn{2}{c}{CIFAR}                   \\ \midrule
Metric  & MTA                                & DAC    & MTA                                & DAC    & MTA                                & DAC    & MTA                                & DAC    \\ \midrule
CFFL    & 85.47 $\pm$ 0.68                   & 100.00 & 78.79 $\pm$ 1.03                   & 100.00 & 89.22 $\pm$ 0.15                   & 100.00 & 65.38 $\pm$ 0.50                   & 100.00 \\
FLTrust & 87.40 $\pm$ 0.71                   & 46.65  & 80.94 $\pm$ 1.26                   & 66.11  & 89.40 $\pm$ 0.20                   & 51.62  & \textbf{70.66} $\pm$ \textbf{0.45} & 39.44  \\
Ours    & \textbf{87.67} $\pm$ \textbf{0.57} & 44.38  & \textbf{83.35} $\pm$ \textbf{1.54} & 65.61  & \textbf{89.88} $\pm$ \textbf{0.15} & 45.27  & 70.28 $\pm$ 0.27                   & 34.89  \\ \bottomrule
\end{tabularx}
\caption{MTA (\%) and DAC (\%) when the \dacquirer has an unbiased root dataset.} \label{tab:even_client}
\end{table*}

\subsubsection{Unbiased Root Dataset}
\label{subsubsec:eval:unbias}
In this segment, we evaluate the scenario where the \dacquirer's root dataset is unbiased. The total number of data samples owned by each \dprovider follows the \textsf{POW} distribution. However, each \dprovider has evenly distributed class labels. The results are shown in Table \ref{tab:even_client}. In general, when the root dataset is reliable, all three methods have better MTAs than the case where the root dataset is biased. \sys achieves slightly better or comparable MTAs compared with other methods with the lowest DACs in all four tasks. 

With the results in Table~\ref{tab:biased_client} and Table~\ref{tab:even_client}, we demonstrate that \first CFFL is slightly more resilient against a biased root dataset than FLTrust. Yet, CFFL introduces consistently high DACs, which is less desirable in data marketplace. \second FLTrust, on the other hand, heavily depends on the root dataset, and can only achieve comparable MTAs with CFFL when the root dataset is unbiased. 
In contrast, \sys produces the best MTAs in nearly all cases  regardless of whether the root dataset is biased or not. Crucially, \sys maintains the lowest DACs when achieving comparable MTAs with the other two methods.


\subsubsection{Robustness Against Various Attacks}
\label{subsubsec:eval:robust}

In this case, we consider the robustness of \sys when facing malicious \dproviders. We compare \sys with both client-driven and server-driven approaches. 
Since we investigate \revision{nearly 700 different combinations of approaches}, attacks, and tasks, we train each combination for a fixed number of 100 epochs in this segment.

First, Figure \ref{fig:trec_untargeted_attack} presents the MTA of each scheme under free-rider attack and sign-randomizing attack on the TREC and CIFAR dataset. The result shows that \sys can defend against the attacks even 80\% of the \dproviders are malicious. 
For the free-rider attack, the MTA of \sys slightly decreases by 2.80\% when the number of faulty \dproviders increases from 30\% to 80\%. For the sign-randomizing attack, the MTA of \sys remains consistent given different numbers of faulty \dproviders. 

\begin{figure}[!t]
    \centering  
    \subfigbottomskip=1pt 
    \subfigcapskip=-5pt 
    \subfigure[Free-rider Attack on TREC]{\label{fig:trec_untargeted_attack_a}
        \includegraphics[width=0.41\linewidth]{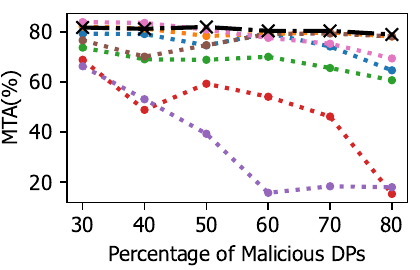}}
    \subfigure[Sign-randomizing attack on TREC]{\label{fig:trec_untargeted_attack_b}
        \includegraphics[width=0.54\linewidth]{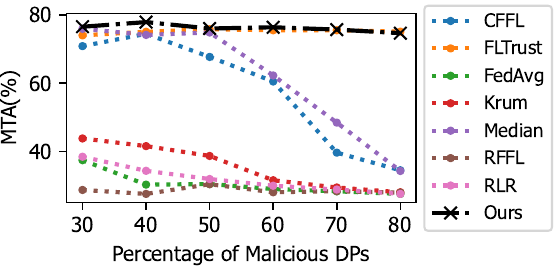}}
    \subfigure[Free-rider Attack on CIFAR]{\label{fig:cifar_untargeted_attack_a}
        \includegraphics[width=0.41\linewidth]{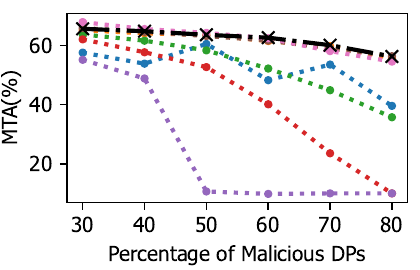}}
    \subfigure[Sign-randomizing attack on CIFAR]{\label{fig:cifar_untargeted_attack_b}
        \includegraphics[width=0.54\linewidth]{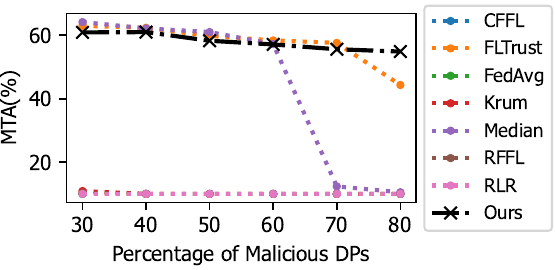}}
    \caption{\revision{The MTA of the global model obtained by different aggregation protocols under untargeted attacks.}}
 \label{fig:trec_untargeted_attack}
\end{figure}

\begin{figure}[!t]
    \centering  
    \subfigbottomskip=1pt 
    \subfigcapskip=-5pt 
    \subfigure[MTA(\%) against Backdoor Attack on TREC]{\label{fig:trec_targeted_attack_a}
        \includegraphics[width=0.41\linewidth]{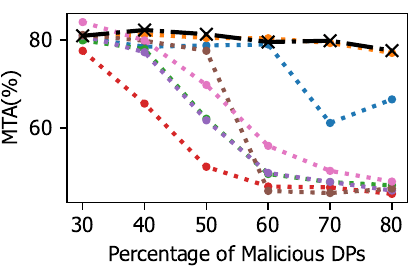}}
    \subfigure[ASR(\%) against Backdoor Attack on TREC]{\label{fig:trec_targeted_attack_b}
        \includegraphics[width=0.54\linewidth]{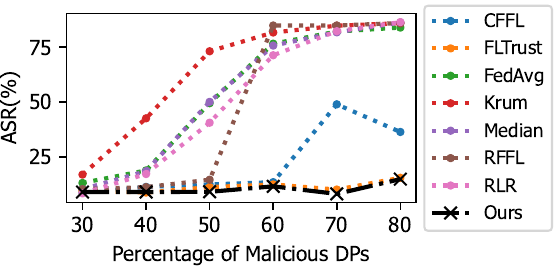}}
    
    \subfigure[MTA(\%) against Label-Flipping Attack on TREC]
    {\label{fig:trec_targeted_attack_c}
        \includegraphics[width=0.41\linewidth]{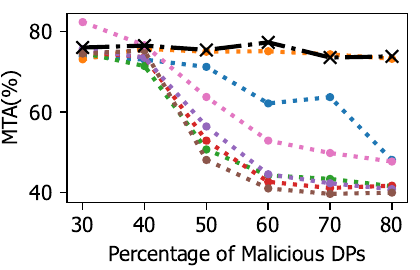}}
    \subfigure[ASR(\%) against Label-Flipping Attack on TREC]
    {\label{fig:trec_targeted_attack_d}
        \includegraphics[width=0.54\linewidth]{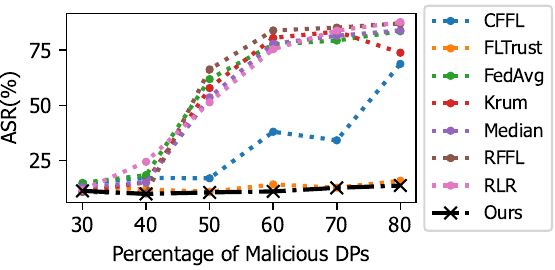}}    
    
    \subfigure[MTA(\%) against Backdoor Attack on CIFAR]{\label{fig:cifar_targeted_attack_a}
        \includegraphics[width=0.41\linewidth]{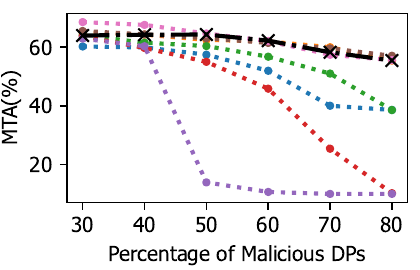}}
    \subfigure[ASR(\%) against Backdoor Attack on CIFAR]{\label{fig:cifar_targeted_attack_b}
        \includegraphics[width=0.54\linewidth]{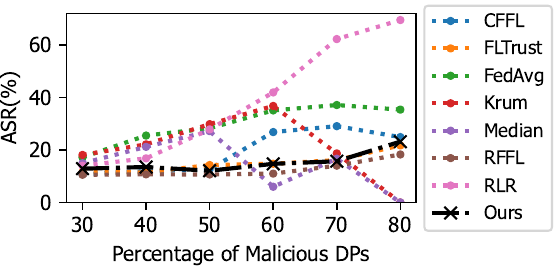}} 
    
    \subfigure[MTA(\%) against Label-Flipping Attack on CIFAR]
    {\label{fig:cifar_targeted_attack_c}
        \includegraphics[width=0.41\linewidth]{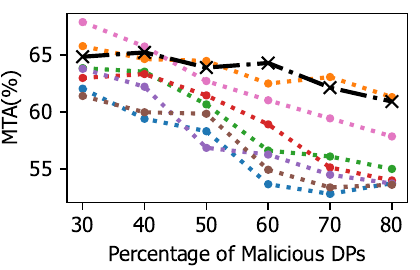}}
    \subfigure[ASR(\%) against Label-Flipping Attack on CIFAR]
    {\label{fig:cifar_targeted_attack_d}
        \includegraphics[width=0.54\linewidth]{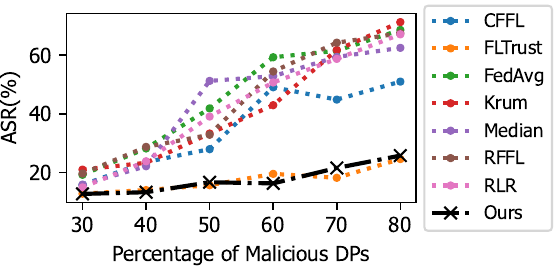}}
    \caption{\revision{The MTA and ASR of the global model obtained by different aggregation protocols under targeted attacks.}}
 \label{fig:trec_targeted_attack}
\end{figure}

Second, we plot the robustness of different aggregation schemes against targeted attacks on the TREC and CIFAR dataset in Figure \ref{fig:trec_targeted_attack}. 
The results show that \sys can achieve the highest MTA and lowest ASR in most cases.  
\revision{When the proportion of attackers is low, such as 30\%, the MTA of RLR~\cite{ozdayi2021defending} is slightly better than other approaches. However, as the proportion of attackers increases, the performance all client-driven methods rapidly deteriorates.} FLTrust has comparable robustness with \sys.  
Note that in Figure \ref{fig:cifar_targeted_attack_b}, the ASR of Krum and Median initially increase, but then decrease to 0. However, the MTAs of both methods also decrease to zero, as shown in Figure \ref{fig:cifar_targeted_attack_a}. This indicates that the global model is not converged for both methods when the percentage of malicious \dproviders is over 60\%.



\begin{figure}[!t]
    \centering  
    \subfigbottomskip=1pt 
    \subfigcapskip=-5pt 
    \subfigure[TREC]{
        \includegraphics[width=0.41\linewidth]{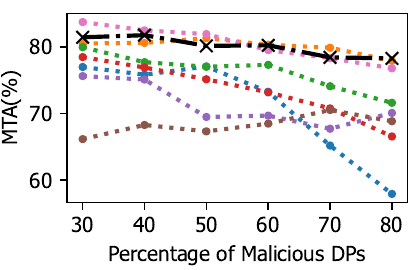}}
    \subfigure[CIFAR]{
        \includegraphics[width=0.54\linewidth]{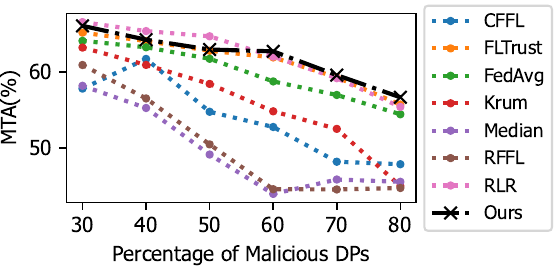}}
    \caption{\revision{The MTA of the global model obtained by different aggregation protocols against Sybil attacks.}}
 \label{fig:sybil_attack}
\end{figure}

Finally, we evaluate the robustness of different schemes against the Sybil attack in Figure \ref{fig:sybil_attack}. The experimental results show \sys and FLTrust have comparable MTAs in most cases. In contrast, the MTA of other methods decrease significantly as the number of Sybil nodes increases. This is because the models submitted by Sybil nodes are similar to each other, so that these schemes cannot accurately eliminate the poisoned local model updates. 

To sum up, CFFL and all the client-driven methods are more vulnerable to faulty \dproviders. For instance, some client-driven aggregation protocols cannot produce a meaningful global model when the number of malicious \dproviders is sufficiently large. 
\sys achieves slightly better robustness than FLTrust, yet reducing roughly 13.91\% DAC on average compared with FLTrust, as shown in Table \ref{tab:dac_robustness}. 

\begin{table}[]
\resizebox{\columnwidth}{!}{
\begin{threeparttable}
\begin{tabular}{lllcccccc}
\hline
Dataset                 & ATK                 & Ratio   & 30\%           & 40\%           & 50\%           & 60\%           & 70\%           & 80\%           \\ \hline
\multirow{10}{*}{TREC}  & \multirow{2}{*}{FR} & FLTrust & 99.22          & 84.38          & \textbf{36.18} & \textbf{31.4}  & 24.22          & 13.68          \\
                        &                     & Ours    & \textbf{87.32} & \textbf{68.32} & 46.58          & 31.54          & \textbf{20.26} & \textbf{12.02} \\ \cline{2-9} 
                        & \multirow{2}{*}{SR} & FLTrust & \textbf{61.64} & \textbf{53.32} & 42.82          & \textbf{33.50} & 26.18          & \textbf{16.06} \\
                        &                     & Ours    & 67.44          & 57.28          & \textbf{40.3}  & 37.98          & \textbf{22.82} & 17.4           \\ \cline{2-9} 
                        & \multirow{2}{*}{BD} & FLTrust & 80.54          & 68.60          & 72.24          & 67.38          & 53.72          & 45.62          \\
                        &                     & Ours    & \textbf{53.76} & \textbf{51.50} & \textbf{46.60} & \textbf{39.60} & \textbf{29.64} & \textbf{41.86} \\ \cline{2-9} 
                        & \multirow{2}{*}{LF} & FLTrust & 83.06          & 81.24          & 65.28          & 70.66          & 58.04          & 53.46          \\
                        &                     & Ours    & \textbf{61.12} & \textbf{52.40} & \textbf{47.92} & \textbf{56.44} & \textbf{30.02} & \textbf{24.96} \\ \cline{2-9} 
                        & \multirow{2}{*}{SY} & FLTrust & 67.20          & 62.64          & 66.72          & 66.98          & 70.40          & 77.92          \\
                        &                     & Ours    & \textbf{53.76} & \textbf{48.10} & \textbf{48.68} & \textbf{44.9}  & \textbf{50.62} & \textbf{60.66} \\ \hline
\multirow{10}{*}{CIFAR} & \multirow{2}{*}{FR} & FLTrust & 72.22          & 63.24          & 51.09          & 48.56          & 35.82          & \textbf{20.71} \\
                        &                     & Ours    & \textbf{68.02} & \textbf{60.36} & \textbf{45.91} & \textbf{46.49} & \textbf{33.02} & 22.13          \\ \cline{2-9} 
                        & \multirow{2}{*}{SR} & FLTrust & \textbf{59.76} & \textbf{53.82} & \textbf{47.24} & \textbf{40.29} & \textbf{33.33} & \textbf{25.93} \\
                        &                     & Ours    & 76.73          & 63.38          & 57.6           & 52.18          & 46.16          & 30.73          \\ \cline{2-9} 
                        & \multirow{2}{*}{BD} & FLTrust & 72.29          & 72.20          & 68.27          & 73.09          & 69.67          & 73.64          \\
                        &                     & Ours    & \textbf{60.93} & \textbf{60.64} & \textbf{54.44} & \textbf{54.56} & \textbf{70.62} & \textbf{49.27} \\ \cline{2-9} 
                        & \multirow{2}{*}{LF} & FLTrust & 74.62          & 74.38          & 72.69          & 72.76          & 70.42          & 66.42          \\
                        &                     & Ours    & \textbf{65.98} & \textbf{60.73} & \textbf{63.07} & \textbf{56.27} & \textbf{54.42} & \textbf{44.58} \\ \cline{2-9} 
                        & \multirow{2}{*}{SY} & FLTrust & 61.29          & 65.38          & 68.42          & 70.2           & 64.38          & 61.33          \\
                        &                     & Ours    & \textbf{56.00} & \textbf{50.53} & \textbf{48.27} & \textbf{44.2}  & \textbf{40.36} & \textbf{39.33} \\ \bottomrule 
\end{tabular}
\begin{tablenotes}
        \footnotesize
        \item[*] In this table, ``ATK'' represents the type of attacks, ``FR'' represents the free-rider attack, ``SR'' represents the sign-randomizing attack, ``BD'' represents the backdoor attack, ``LF'' represents the label-flipping attack, and ``SY'' represents the Sybil attack.
      \end{tablenotes}
    \end{threeparttable}
}
\caption{The DAC (\%) comparison between FLTrust and \sys in the robustness experiments. Results for CFFL are omitted since they are always 100.}\label{tab:dac_robustness}
\end{table}

\subsubsection{Accuracy Loss by Quantization}
\label{subsubsec:eval:quantization}

To enable verifiable data transaction in \sys, the \dacquirer needs to first quantize the prior model $W_g^{t-1}$ and local model updates $U^t$, perform the aggregation to obtain quantized global model $\mathbb{W}_g^t$, and then de-quantize the model to obtain a floating-point model $W_g^t$. 
In this segment, we study the impact of quantization on model accuracy. We evaluate the scenario where the \dacquirer's root dataset is unbiased. The total number of data samples and the type of labels owned by each \dprovider follows the \textsf{POW} and \textsf{IID} distribution, respectively. 
The results are summarized in Table~\ref{tab:quantization}. 
The $\Delta QT$ represents the MTA and F1 loss due to quantization (\ie the accuracy difference between $W_g^t$ and $\mathbb{W}_g^t$).
The results indicate that the quantization operations introduce negligible accuracy losses.  

\begin{table}[]\small
\begin{tabular}{@{}lcccc@{}}
\toprule
\multirow{2}{*}{Dataset} & \multicolumn{2}{c}{MTA}                                    & \multicolumn{2}{c}{F1}                                     \\ \cmidrule(l){2-5} 
                         & $QT$             & $\Delta QT$                             & $QT$             & $\Delta QT$                             \\ \midrule
TREC                     & 88.60 $\pm$ 0.28 & -0.93 \textcolor{darkgreen}{$\uparrow$} & 87.58 $\pm$ 0.37 & -0.65 \textcolor{darkgreen}{$\uparrow$} \\
AGNEWS                   & 82.61 $\pm$ 0.65 & \;0.74 \textcolor{darkred}{$\downarrow$}  & 82.54 $\pm$ 0.65 & \;0.64 \textcolor{darkred}{$\downarrow$}  \\
FMNIST                   & 90.07 $\pm$ 0.07 & -0.20 \textcolor{darkgreen}{$\uparrow$} & 90.03 $\pm$ 0.07 & -0.20 \textcolor{darkgreen}{$\uparrow$} \\
CIFAR                    & 69.74 $\pm$ 0.62 & \;0.54 \textcolor{darkred}{$\downarrow$}  & 69.74 $\pm$ 0.69 & \;0.53 \textcolor{darkred}{$\downarrow$}  \\ \bottomrule
\end{tabular}\caption{The MTA(\%) and F1(\%) loss in  quantization.}\label{tab:quantization} 
\end{table}

\subsubsection{System-level Overhead}
\label{subsubsec:eval:efficiency}
In this segment, we study the system-level overhead of \first, including the cryptography overhead in our quality-aware model evaluation protocol; \second the time and gas cost of the functions in the trading smart contract.

\begin{figure}[]
\centering
\begin{minipage}[t]{0.335\columnwidth}
\centering
\includegraphics[width=\linewidth]{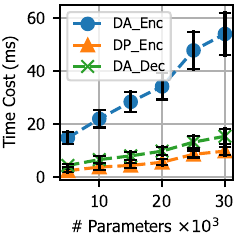}
\caption{The homomorphic ENC/DEC times.}
\label{fig:encryption}
\end{minipage}
\begin{minipage}[t]{0.63\columnwidth}
\centering
\includegraphics[width=\linewidth]{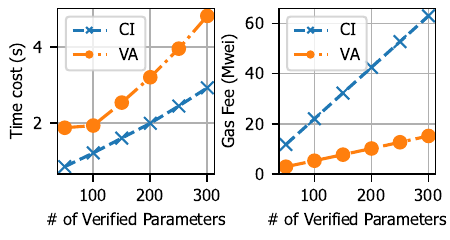}
\caption{The gas cost and execution times for verifying different numbers of parameters.}
\label{fig:verify}
\end{minipage}
\end{figure}

In Figure \ref{fig:encryption}, we plot the overhead of homomorphic encryption and decryption operations. 
In the experiment, we set 4096 slots per batch for the CKKS algorithm. The encryption time of the \dacquirer is longer because the \dacquirer needs to perform homomorphic encryption, while \dproviders only needs to complete homomorphic additions. 
Overall, the extra latency introduced by homomorphic operations is small. 
\begin{table}[]
\resizebox{\columnwidth}{!}{
\begin{threeparttable}
\begin{tabular}{@{}lccccccc@{}}
\toprule
Phrase    & \multicolumn{4}{c}{Prepare}         & \multicolumn{2}{c}{Verify} & Any  \\ \midrule
Function  & NE & \deposit & CM & \prepare & CR(\dacquirer)& CR(\dprovider) & RE   \\ \midrule
Gas (wei) & 163076 & 46643   & 127731 & 222018  & 38036        & 42632       & -    \\
Time (ms) & 92.0   & 73.6    & 80.6   & 83.4    & 64.2         & 70.6        & 74.8 \\ \bottomrule
\end{tabular}
\begin{tablenotes}
        \footnotesize
        \item[*] In this table, ``NE'' represents \newepoch, ``CM'' represents \register,``CR'' represents \claim, and ``RE'' represents \readepoch.
      \end{tablenotes}
    \end{threeparttable}
}
\caption{The gas costs and execution times of the functions in our trading smart contract.}\label{tab:smart_contract}
\end{table}

\revision{Further, we report the gas cost and latency for executing different functions in our trading smart contract in \S~\ref{subsec:payment}.}
In the \preparephase, \newepoch \xspace initializes training epochs, \deposit \xspace enables the \dacquirer to deposit rewards and penalties. \register \xspace allows the \dproviders to commit local models, and \prepare \xspace records rewards and selects verification parameters. 
In the \verifyphase, \commit \xspace (abbreviated as ``CI'') allows the \dacquirer to commit the public inputs for aggregation verification, and \vrf \xspace (abbreviated as ``VA'') verifies the integrity of aggregation. Both the \dacquirer and \dproviders can use \claim \xspace to claim rewards. Finally, \readepoch \xspace is a convenience handle to return detailed epoch information.
\iftechreport
The detailed implementations are deferred to \S~\ref{subsec:pseudocode}}. 
\fi
Overall, none of these functions consumes more than $0.25 \times 10^6$ wei, which costs less than $0.0025$ US dollars at the time of this writing. 

In comparison to Omnilytics~\cite{liang2021omnilytics}, which directly aggregates local models via a smart contract, the gas cost in \sys is at least 1000 times smaller, when enabling approximately 8 times more participants to collectively train models with ${\sim}100$ times more parameters than the model trained in Omnilytics. In fact, the gas cost in \sys is independent of the model size and the model evaluation method, since \sys only verifies a fixed number (a system setting) of randomly sampled model parameters.
In addition, the execution time of each function is less than $0.1$ seconds. In Figure~\ref{fig:verify}, we also report the gas cost and execution time when \sys has different system settings that verify different numbers of model parameters.

%% file: discussion.tex
\subsection{Deep Dive}
\label{subsec:deep_dive}

In this section, we further investigate several key design choices of \sys and outline some future work.

\subsubsection{Tradeoff Between Inclusiveness and Robustness}\label{subsubsec:tradeoff_analysis}

In \S~\ref{subsubsec:key_observation}, we presented the key observation that existing approaches exhibit a fundamental tradeoff between inclusiveness and robustness when aggregating local models submitted by the \dproviders. In this segment, we further analyze this tradeoff by tuning the key parameters in the aggregation algorithms of both FLTrust~\cite{cao2021fltrust} and Krum~\cite{blanchard2017machine}. 
Specifically, since FLTrust selects local models based on the cosine similarities between them and the self-computed model update trained on the \dacquirer's root dataset, the key system parameter that dictates the aggregation in FLTrust is the quality of the root dataset. We quantify the quality as \emph{unbiasness ratio}, which represents the percentage of class labels that predominates the \dacquirer's root dataset. For instance, FLTrust-1/3 represents the case where the root dataset contains 1/3 of the class labels. In Krum, the key tunable parameter $f$ is the proportion of Byzantine \dproviders defined in Krum's problem formulation, which directly determines the number of local models selected for aggregation in each epoch. We evaluate four Krum settings in this part (for instance, Krum-50 represents the setting where the Krum algorithm is supposed to tolerate 50\% Byzantine \dproviders).  
We consider a group of heterogeneous \dproviders, where 30\% of them hold high-quality data (evenly distributed across all types of labels), 30\% of them hold biased datasets in which the class labels are dominated by half of the randomly selected labels, and 40\% of them are malicious. We experiment both the targeted and untargeted attacks on the TREC and FMNIST tasks.

\begin{figure}[!t]
    \centering  
    \subfigbottomskip=1pt 
    \subfigcapskip=-5pt 
    \subfigure[Backdoor Attack on TREC]{\label{fig:tradeoff_backdoor}
        \includegraphics[width=0.48\linewidth]{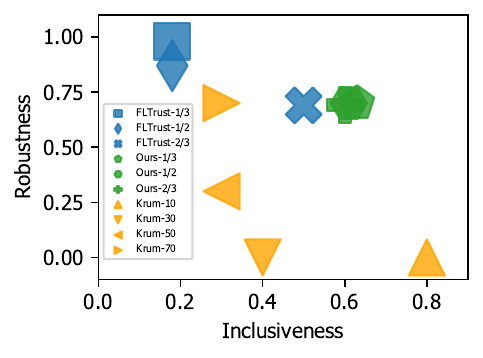}}
    \subfigure[Sign-Randomizing Attack on TREC]{\label{fig:tradeoff_sign}
        \includegraphics[width=0.48\linewidth]{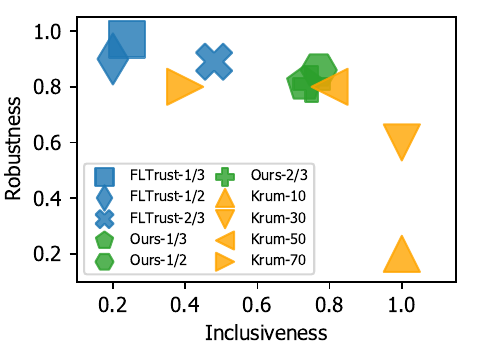}}
    
    \subfigure[Label-Flipping Attack on FMNIST]{\label{fig:tradeoff_flipping}
        \includegraphics[width=0.48\linewidth]{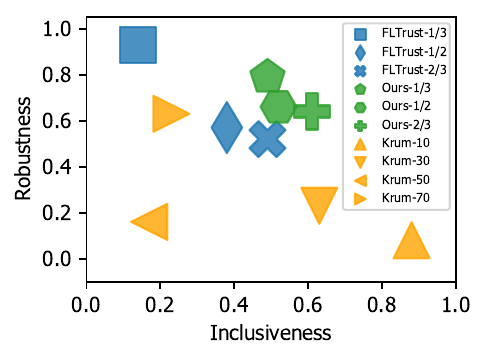}}
    \subfigure[Free-Rider Attack on FMNIST]{\label{fig:tradeoff_free}
        \includegraphics[width=0.48\linewidth]{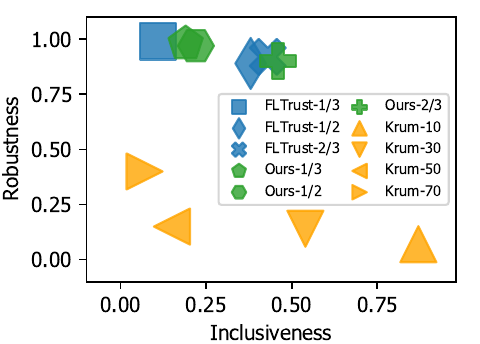}}    
        
    \caption{Deep dive into the tradeoff between inclusiveness and robustness in various settings.}
 \label{fig:tradeoff_analysis}
\end{figure}

The results are plotted in Figure~\ref{fig:tradeoff_analysis}, where inclusiveness is the percentage of benign \dproviders whose local models are selected for aggregation, and robustness is quantified as the percentage of malicious \dproviders that are excluded for aggregation. We collect these two metrics in each training epoch and report the average values. In general, FLTrust can achieve reasonably good inclusiveness only if the unbiasness ratio of the \dacquirer's root dataset is sufficiently large (\eg reaching 2/3). Krum, which is significantly impacted by its  Byzantine tolerance threshold $f$, tends to have high robustness at the expense of inclusiveness.  
In contrast, \sys achieves the best tradeoff between inclusiveness and robustness in all cases. As a result, \sys consistently achieves high model performance regardless of the parameter settings, as summarized in Table~\ref{tab:tradeoff_analysis}. However, FLTrust begins to match \sys only when the root dataset is sufficiently good. Krum fails to perform consistently across all tasks, regardless of the parameter choices. For instance, although Krum-50 achieves a good MTA on TREC against the sign-randomizing attack, it has a significantly high ASR against the backdoor attack.

\begin{table}[]\small
\begin{threeparttable}
\begin{tabular}{|l|lcc|ccc|}
\hline
Dataset     & \multicolumn{3}{c|}{TREC}                  & \multicolumn{3}{c|}{FMNIST}                \\ \hline
Attack      & \multicolumn{2}{c|}{BD}            & SR    & \multicolumn{2}{c|}{LF}            & FR    \\ \hline
Metric      & MTA   & \multicolumn{1}{c|}{ASR}   & MTA   & MTA   & \multicolumn{1}{c|}{ASR}   & MTA   \\ \hline
FLTrust-1/3 & 48.60 & \multicolumn{1}{c|}{12.02} & 57.08 & 74.71 & \multicolumn{1}{c|}{18.96} & 70.71 \\
FLTrust-1/2 & 58.84 & \multicolumn{1}{c|}{9.01}  & 57.48 & 87.11 & \multicolumn{1}{c|}{4.69}  & 86.52 \\
FLTrust-2/3 & 81.53 & \multicolumn{1}{c|}{9.43}  & 83.16 & 87.97 & \multicolumn{1}{c|}{5.37}  & 86.38 \\ \hline
Krum-10     & 70.11 & \multicolumn{1}{c|}{41.58} & 32.08 & 88.43 & \multicolumn{1}{c|}{5.12}  & 87.69 \\
Krum-30     & 61.64 & \multicolumn{1}{c|}{68.27} & 38.24 & 86.44 & \multicolumn{1}{c|}{13.43} & 86.60 \\
Krum-50     & 64.16 & \multicolumn{1}{c|}{67.04} & \textbf{85.08} & 80.64 & \multicolumn{1}{c|}{32.20} & 79.16 \\
Krum-70     & 72.64 & \multicolumn{1}{c|}{36.23} & 78.64 & 85.97 & \multicolumn{1}{c|}{4.95}  & 8.19  \\ \hline
Ours-1/3    & 83.71 & \multicolumn{1}{c|}{7.75}  & 76.36 & 87.94 & \multicolumn{1}{c|}{\textbf{3.41}}  & 81.19 \\
Ours-1/2    & 83.38 & \multicolumn{1}{c|}{8.43}  & 75.36 & 88.23 & \multicolumn{1}{c|}{4.48}  & 84.34 \\
Ours-2/3    & \textbf{84.08} & \multicolumn{1}{c|}{\textbf{7.55}}  & 79.96 & \textbf{88.65} & \multicolumn{1}{c|}{3.94}  & \textbf{88.18} \\ \hline
\end{tabular}
\begin{tablenotes}
        \footnotesize
        \item[*] In this table, ``BD'' represents the backdoor attack, ``SR'' represents the sign-randomizing attack, ``LF'' represents the label-flipping attack, and ``FR'' represents the free-rider attack.
      \end{tablenotes}
    \end{threeparttable}
\caption{The MTA (\%) and ASR (\%) in the inclusiveness-robustness tradeoff experiments.}
\label{tab:tradeoff_analysis}
\end{table}


\begin{table}[]\small
\begin{tabular}{@{}llcccc@{}}
\toprule
Attack                & Scenarios      & TREC  & AGNEWS & FMNIST & CIFAR  \\ \midrule
\multirow{3}{*}{None} & Type-I Biased  & 80.00 & 70.00  & 98.80  & 95.67  \\
                      & Type-II Biased & 88.33 & 98.67  & 100.00 & 100.00 \\
                      & Unbiased       & 95.67 & 100.00 & 99.67  & 99.67  \\ \midrule
\multirow{3}{*}{BD}   & Type-I Biased  & 65.33 & 58.00  & 96.67  & 95.00  \\
                      & Type-II Biased & 80.33 & 93.00  & 95.33  & 100.00 \\
                      & Unbiased       & 79.00 & 84.00  & 100.00 & 100.00 \\ \midrule
\multirow{3}{*}{SR}   & Type-I Biased  & 64.67 & 62.67  & 98.67  & 99.00  \\
                      & Type-II Biased & 96.00 & 88.67  & 99.00  & 89.50  \\
                      & Unbiased       & 95.67 & 99.67  & 99.33  & 91.50  \\ \bottomrule
\end{tabular}
\begin{tablenotes}
        \footnotesize
        \item[*] ``BD'' and ``SR'' represent the backdoor and sign-randomizing attack, respectively.
      \end{tablenotes}
\caption{The probability of selecting correct baselines by our dynamic baseline adjustment algorithm.}\label{tab:baseline_correctness}
\end{table}

\subsubsection{Analysis of Dynamic Baseline Adjustment}\label{subsubsec:aggregation_analysis}

\parab{Quantitative Results.} We first quantitatively analyze the accuracy of our Dynamic Baseline Adjustment algorithm (detailed in \S~\ref{subsec:dynamic_selection}) to demonstrate its robustness in various data distributions. We consider a challenging setting for the \dproviders, where 30\% of them hold high-quality data and 30\% of them hold biased datasets. The  remaining 40\% of \dproviders have three different settings: \first holding evenly-distributed data, \second maliciously engaging the backdoor attack (one type of targeted attack), and \third maliciously engaging the sign-randomizing attack (one type of untargeted attack).
In terms of the root dataset held by the \dacquirer, we consider the following three scenarios:

\begin{itemize}[leftmargin=15pt]
    \item \textbf{Type-I Bias.} The root dataset of \dacquirer is dominated by half of the class labels. The data distributions of the biased \dproviders are similar to the distributions of the \dacquirer's root dataset. 
    \item \textbf{Type-II Bias.} The root dataset of \dacquirer is dominated by half of the class labels. However, the data distributions of the biased \dproviders are random and independent of the \dacquirer's data distributions.
    \item \textbf{Unbiased.} The root dataset of \dacquirer is evenly distributed.
\end{itemize}

We analyze the probability of selecting correct baselines during the  training process. A baseline is correct if the cosine similarity between the baseline and the \emph{ground truth model update} is strictly positive. The ground truth update is an ideal update obtained directly on high-quality data, which can be considered as \emph{a hypothetical scenario} where the \dacquirer possesses sufficient high-quality data and can train the model all by itself.

The results are reported in Table~\ref{tab:baseline_correctness}. The Type-I Bias is arguably the most challenging scenario because the \dacquirer and the biased \dproviders are \emph{similarly biased}. Therefore, the \dacquirer is prone to be misled by these biased \dproviders, resulting in possible incorrect baseline selections (note that all our experiments in \S~\ref{subsec:evaluation_results} considered the Type-I Bias). 
Nonetheless, our method still achieves reasonably accurate baseline selections, up to 99\% accuracy in the training epochs of the CIFRA task. We also observed that the selection accuracies are affected by the total number of class labels. Specifically, TREC and AGNEWS have 6 and 4 class labels, respectively, while FMNIST and CIFAR both have 10 class labels. When the total number of class labels is smaller, the data diversity experienced by the biased \dacquirer and \dproviders is even lower (for instance, they only see 2 labels in the AGNEWS task). This results in a relatively higher probability of selecting incorrect baselines.   
In the Type-II Bias scenario, where the data distributions of the biased \dproviders and the \dacquirer are not correlated, the probabilities of selecting correct baselines are higher than the Type-I Bias scenario, even for the TREC and AGNEWS task. In fact, the selection accuracies in the Type-II Bias scenario are already comparable to the case where the \dacquirer's root dataset is unbiased, achieving over 90.0\% baseline selection accuracy in most settings.

\parab{Attack Discussion.} To attack our model aggregation protocol, an adversary must carefully design local models that can be selected as the baseline. However, there are several challenges to crafting such models. First, the adversary has no access to the plaintext models submitted by other \dproviders throughout the model aggregation and transaction. Instead, the adversary only observes the commitments of these local models, which do not reveal the actual local models. Additionally, because the \dacquirer evaluates all local models offline using the private model evaluation algorithm with its private root dataset, it is difficult for the adversary to predict what types of local models will receive higher Kappa scores in the root dataset. Finally, the \dacquirer discloses the \dprovider that is selected as the baseline for the current epoch only after all \dproviders have committed their models. At this point, the adversary cannot modify its committed local model, even if it colludes with the selected \dprovider. In summary, because \sys enables complete offline and private local model evaluation and aggregation by the \dacquirer, attacking our aggregation protocol is significantly more difficult than  existing blockchain-based approaches (\eg Omnilytics~\cite{liang2021omnilytics}, FPPDL~\cite{lyu2020towards}) that require the aggregation protocol to be publicly observable on the blockchain.

\parab{Future Work on Model Evaluation and Aggregation Protocols.}
Several studies \cite{allen2020byzantine, karimireddy2021learning} have shown that historical information can be used to identify Byzantine \dproviders. In our future work, we intend to leverage this insight by taking into account \dproviders' historical reputations and incorporate momentum into our local model evaluation algorithm. This approach may further reduce the reliance on the \dacquirer's root dataset and compensate for errors in baseline selections, particularly in the Type-I Bias scenario.

%% file: 7.related.tex
\section{Related Work}
\label{sec:related}


\iftechreport

\parab{Robust FL Constructions.}
In \S~\ref{sec:evaluation}, we compared \sys with five client-driven methods (FedAvg~\cite{mcmahan2017communication}, RFFL~\cite{xu2020reputation}, Krum~\cite{blanchard2017machine}, Median~\cite{yin2018byzantine}, and RLR~\cite{chen2021robust}), and two server-driven methods (FLTrust~\cite{cao2021fltrust} and CFFL~\cite{lyu2020collaborative}). We summarize their core constructions here.
McMahan et al.~\cite{mcmahan2017communication} proposed FedAvg, in which the aggregation weights of the \dproviders are proportional to their data volumes. Yin et al. \cite{yin2018byzantine} showed that aggregating local models using the median model update or trimmed mean of all model updates can be effective against Byzantine attacks. Yin et al.~\cite{blanchard2017machine} proposed the Krum algorithm to only choose the $n{-}f{-2}$ closest vectors from $n$ submitted vectors (assuming the algorithm can tolerate up to $f$ Byzantine clients among all $n$ clients) to aggregate the global model. Xu et al.~\cite{xu2020reputation} designed the RFFL method, which first aggregates the global model, then calculates the similarity between each local model and the global model, and uses it as the reputation of the participants for the aggregation in the next epoch.
\revision{Ozdayi et al.~\cite{ozdayi2021defending} proposed RLR, a lightweight aggregation method. To defend against the backdoor attack, RLR carefully adjusts the aggregation server’s learning rate, per dimension and per round, based on the sign information of the updates from clients.}
Lyu et al.~\cite{lyu2020collaborative} proposed CFFL, a method that validates the accuracy of data provider models based on local datasets of the \dacquirer, and quantifies the collaborative fairness of each \dprovider via the correlation coefficient between participant contributions and participant rewards. In aggregation, CFFL redefines the weight of each model based on its test accuracy and historical contributions. 
In FLTrust~\cite{chen2021robust}, the \dacquirer is required to hold a clean local dataset to evaluate clients' local updates. It compares the similarity in direction and size between the server updates and the local model updates for each participant and uses the similarity as the weight to aggregate the global model. 
\fi

\parab{Verifiable Protocols for Real-world Systems.}
The enthusiasm for Web 3.0~\cite{liu2021make} drives a growing number of literature on empowering or even transforming real-world systems via verifiable (or trust-free) protocols, such as verifiable cloud computing (\eg \cite{dong2017betrayal,li2022verifying}), 
decentralized digital good exchanges (\eg \cite{dziembowski2018fairswap}), smart contract based legal sector transformation~\cite{fang2023isyn}, and various proposals to improve the Blockchain systems themselves (such as  interoperability (\eg \cite{zamyatin2019xclaim,liu2019hyperservice,xie2022zkbridge}) and private smart contracting~\cite{kosba2016hawk, cheng2019ekiden}). 
Overall, practicability and deployability are two the primary challenges for designing verifiable protocols to power real-world systems. Thus, the verifiable transaction protocol in \sys only proves the critical computations that are necessary and sufficient to ensure fair billing. It does not verify the entire FL training process, which would otherwise impose unacceptably high overhead. 

\parab{Data Pricing.}
Prior works have studied data pricing. For instance, proposals~\cite{jia2019towards, ghorbani2019data} evaluate data value based on Shapley Value using a game theoretic approach, which typically requires access to the full datasets. 
Some other literature (\eg \cite{balazinska2011data,  koutris2013toward}) propose a pricing framework for relational queries. Data pricing for FL-based marketplace is part of our future work.

%% file: 8.conclusion.tex
\section{Conclusion}

In this paper, we propose \sys, a novel FL architecture that is specifically designed to enable a utility-driven data marketplace. Benefiting from the quality-aware model evaluation protocol, \sys can eliminate the tradeoff between inclusiveness and robustness when selecting desired \dproviders. 
Further, \sys designs a verifiable transaction protocol that enables the \dacquirer to prove that it faithfully aggregates model using the committed aggregations weights, enabling fair 
trading between the \dacquirer and \dproviders. We implemented a prototype of \sys and extensively evaluated it on four datasets. The experimental results demonstrate the accuracy, robustness and efficiency of \sys. 

%% file: appendix.tex
\appendix
\section{Appendix}
\label{sec:appendix}

\subsection{Pseudocode of Trading Smart Contract}
\label{subsec:pseudocode}
The pseudocode of our trading smart contract is presented in Algorithm \ref{alg:pseudocode}. 

\begin{algorithm*}[htb]
\caption{The Pseudocode of Our Trading Smart Contract}
\label{alg:pseudocode} 
    \begin{multicols}{2}
    \small
    \struct \gone \{\xspace \uuint X,\xspace \uuint Y \}\\
    \struct \gtwo \{\uuint[2] X,\xspace \uuint[2] Y \}\\
    \struct \textsf{PF} \{\xspace \gone $a$, $c$, \gtwo $b$ \}{\color{comment} \# Proof} \\
    
    \struct \textsf{VK} \{ \gone $\alpha$, \gone[] $\gamma_{abc}$, \gtwo $\beta$, $\gamma$, $\delta$ \} \\

    \struct \textsf{EP} \{ {\color{comment} \# Epoch}\\
    \quad \uuint depositDP, {\color{comment} \# The deposit fund for \dproviders.} \\
    \quad \uuint depositDA,{\color{comment} \# The deposit fund for \dacquirer.} \\
    \quad \uuint delay, {\color{comment} \# The maximum delay for registering in this epoch.} \\
    \quad \uuint ts, {\color{comment} \# The timestamp to initiate this epoch.} \\
    \quad \uuint[] samples, {\color{comment} \# The verified parameters.}\\
    \quad \mmapping (\uuint => \uuint) amount,{\color{comment} \# The reward amount for each \dprovider.}\\
    \quad \mmapping (\uuint => \sstring) model,{\color{comment} \# The model info from each \dprovider.}\\
    \quad \mmapping (\sstring =>\uuint[]) inputs, {\color{comment} \# The public inputs in verification.}\\
    \quad \bbool isRegister, isPrepared, isVerified, isFailed \} \\
    \init \train $:= \emptyset$, dataAcquirer $:=$ \nnull\\
    \init numEP $:= 0$, numDP $:= 0$. {\color{comment} \# Number of epoch and \dprovider.}\\
    \init \mmapping (\aaddr => \uuint) dataProvider $:= \emptyset$\\
    
    \constructor()\bco dataAcquirer $:=$ \msg.sender\\

    \modifier \onlyowner \bco \req \msg.sender = dataAcquirer
    

    \func \register($\epoch$, hash, sig)\public \bco \\
    \quad \req \epoch < numEP \textbf{and} \train[\epoch].isRegister = \fal\\
    \quad \textsf{EP} e $:=$ \train[\epoch]\\
    \quad \textbf{if} ((\timestamp - e.ts) > e.delay) \bco \\ 
    \quad \quad e.isRegister $:=$ \tru, \abort {\color{comment} \# register phase is done}\\
    \quad \textbf{else} \bco \\ 
    \quad \quad \textbf{if} \msg.sender \textbf{not in} dataAcquirer.keys \bco \\
    \quad \quad \quad dataProvider[\msg.sender] $:=$ numDP, numDP ++\\
    \quad \quad e.model[dataProvider[\msg.sender]] $:= $ hash, sig\\
    
    \func \newepoch($T$) \public \payable \onlyowner \bco\\
    \quad \textbf{if} numEP!=0 \bco {\color{comment} \# Only when the previous epoch is verified}\\
    \quad \quad \req(\train[numEP-1].isVerified = \tru)\\
    \quad \textsf{EP} e $:=$ \train.push()\\
    \quad e.depositDA, e.depositDP $:=$ \msg.value/2\\
    \quad e.isRegister,e.isPrepared,e.isVerified,e.isFailed $:=$ \fal\\
    \quad e.ts $:=$ \timestamp, e.delay $:=$ $T$\\
    \quad \emit \textsf{NewEpoch}(numEP,e.ts + e.delay)\\
    \quad numEP ++\\

    \func \readepoch($\epoch$)\public \bco \\
    \quad \req \epoch < numEP \\
    \quad return \train[\epoch]

    \func \prepare(\epoch, addr, amount, num) \public \onlyowner \bco \\
    \quad \req \epoch < numEP and \train[\epoch].isPrepared = \fal \\
    \quad \req \len(addr) = \len(amount)\\
    \quad \textsf{EP} e $:=$ \train[\epoch]\\
    \quad \textbf{if} ((\timestamp - e.ts) > e.delay) \bco e.isRegister $:=$ \tru\\
    \quad \textbf{else} \bco \abort {\color{comment} \# register phase in progress}\\
    \quad \req \len(addr) = \len(e.model) \textbf{and} sum(num) = sum(e.amount)\\
    \quad \textbf{for} $i$ \textbf{in} range(\len(addr)) \bco e.amount[addr[i]] $:=$ amount[i]\\
    \quad e.samples,$\pi_{\textsf{vdf}} :=$ \vdf(num) {\color{comment} \# generate ramdom samples}\\
    \quad e.isPrepared $:=$ \tru \\
    \quad \emit \textsf{EpochPrepareEvent}(\epoch)\\

    \func \commit($\epoch, vk, \pi^t_{\textsf{agg}}, \mathbb{X}^{t,c}$)\public \onlyowner \bco \\
    \quad \req \epoch < numEP \textbf{and} \train[\epoch].isPrepared = \tru \\
    \quad \req \train[\epoch].isVerified = \fal \\
    \quad \train[\epoch].inputs = $vk, \pi^t_{\textsf{agg}}, \mathbb{X}^{t,c}$ \\
    \quad \emit \textsf{CommitPublicInputs}(\epoch)\\
    
    \func \vrf($\epoch$) \public \bco \\
    \quad \req \epoch < numEP\\
    \quad \textsf{EP} e $:=$ \train[\epoch]\\
    \quad \req e.isPrepared = \tru \textbf{and} e.isVerified = \fal\\
    \quad $vk,\mathbb{X}^{t,c},\pi^t_{\textsf{agg}} =$ e.inputs\\
    \quad $v := \verify(vk, \mathbb{X}^{t,c}, \pi^t_{\textsf{agg}})$\\
    \quad e.isVerified $:=$ \tru \\
    \quad \textbf{if} $v$ = \fal \bco \\
    \quad \quad e.isFailed $:=$ \tru\\
    \quad \quad \textbf{for} id \textbf{in} e.amount.keys \bco \\
    \quad \quad \quad e.amount[id] += e.depositDA/\len(e.amount.keys)\\
    \quad \quad e.depositDA $:=$ 0\\
    \quad \emit \textsf{EpochVerified}(\epoch,$v$)\\

    \func \verify($vk, x, \pi$) \private \bco \\
    \quad \req \len($x$) $+ 1 = \len($vk.$\gamma_{abc}$)\\
    \quad \gone tmp $:=$ \gone$(0, 0)$\\
    \quad \textbf{for} $i$ \textbf{in} range(\len($x$)) \bco \\
    \quad \quad \req (x[i] < p) {\color{comment} \# $p$ the large prime number in \textsf{G2} group.} \\
    \quad \quad tmp $:= $ \add(tmp, \smul(vk.$\gamma_{abc}$[i + 1], x[i]))\\
    \quad tmp $:=$ \add(tmp, vk.$gamma_{abc}$[0])\\
    \quad $p_1$ := $\pi$.a, \negate(tmp),\negate($\pi$.c),\negate(vk.$\alpha$)\\
    \quad $p_2$ := $\pi$.b, vk.$\gamma$,vk.$\delta$,vk.$\beta$\\
    \quad \return \pair($p_1$,$p_2$)\\

    \func \deposit($\epoch$) \public \payable \onlyowner \bco \\
    \quad \req \epoch < numEP \textbf{and} \train[epoch].isRegister = \fal \\
    \quad \textsf{EP} e $:=$ \train[\epoch]\\
    \quad \textbf{if} ((\timestamp - e.ts) > e.delay) \bco \\ 
    \quad \quad e.isRegister $:=$ \tru, \abort {\color{comment} \# register phase is done}\\
    \quad \textbf{else} \bco \\ 
    \quad \quad e.depositDA += \msg.value/2;\\
    \quad \quad e.depositDP += \msg.value/2;\\
    \quad \quad \emit \textsf{EpochDepositEvent}(\epoch,e.depositDA)\\
    
    \func \claim(\epoch) \public \payable \bco \\
    \quad \req \epoch < numEP \textbf{and} \train[epoch].isVerified = \tru\\
    \quad \textsf{EP} e $:=$ \train[\epoch]\\
    \quad ID $:=$ dataProvider[\msg.sender]\\
    \quad \req \msg.sender = dataAcquirer \textbf{or} ID \textbf{in} e.model.keys\\
    \quad \textbf{if} \msg.sender = dataAcquirer \bco \\
    \quad \quad amount $:=$ e.depositDA \\
    \quad \quad \textbf{if} (\msg.sender.send(amount)) \bco e.depositDA $:=$ 0 \\
    \quad \textbf{else} \bco\\
    \quad \quad  amount $:=$ e.amount[ID]\\
    \quad \quad \textbf{if} (\msg.sender.send(amount)) \bco e.amount[ID] $:=$ 0\\
    \end{multicols}
\end{algorithm*}

\parab{\textbf{Data Structure.}} In smart contracts, we have designed three data structures: Verification Key (\textsf{VK}), Proof (\textsf{PF}), and Training Epochs (\textsf{EP}). To achieve verifiable computation in smart contracts, we utilize the ALT\_BN128 elliptic curve, which is supported in Ethereum EIP-197~\cite{eip197}. 

The \textsf{EP} data structure stores information related to a training epoch. It includes fields such as $depositDP$ and $depositDA$, representing the deposit funds for the \dprovider and \dacquirer, respectively. Other fields include $delay$, which specifies the delay time for registration, and $ts$, representing the timestamp of when the epoch was initiated. The $samples$ field is an array of verified parameters, and the $amount$ and $model$ fields are mappings representing the tokens for each \dprovider and model information from them, respectively. The boolean fields $isRegister$, $isPrepared$,  and $isVerified$ indicate whether the register phase, preparation phase, and verification phase are completed, respectively. The $isFailed$ identifier presents the verification result.

\parab{\textbf{Training setup.}} At the beginning of the model training, the smart contract initializes the training process $\train$ as an empty set. As the training proceeds, the $\train$ set stores the training data for each epoch in an \textsf{EP} data structure. The $\train$ set can be used for source tracing and supervision of transactions. The smart contract also records the number of completed training epochs $numEpoch$ and the number of registered \dproviders $numDP$. The \dacquirer is responsible for deploying the contract, so $\textsf{dataAcquirer}$ is set to the address of the message sender in the $\constructor$. We use the $\onlyowner$ keyword to restrict certain functions to be called only by the \dacquirer. 

\parab{\textbf{Epoch Initialization.}} The \dacquirer can use the function $\newepoch$ to initiate a new epoch. It creates a new $\textsf{EP}$ structure, adds it to $\train$, and deposits the initial funds through $\msg.\textsf{value}$. If it is not the first epoch, the smart contract will check whether the previous epoch has been verified. This is to ensure that the \dacquirer cannot start a new epoch until the fund allocation for the previous epoch is complete. In addition, when a new epoch is created successfully, the model registration time for the epoch is set to $\textsf{delay}$, and the smart contract broadcasts an event to notify all \dproviders. Also, \dproviders can use $\readepoch$ to get the $\textsf{EP}$ to check the details of the epoch.

\parab{\textbf{Deposit Reward and Penalty.}} The \dacquirer can use the $\newepoch$ and $\deposit$ functions to deposit the funds in the epoch, including the rewards for \dproviders proportional to their weights in $\mathbb{K}^t$ and the penalty if it cannot later provide a correct proof. When the epoch is initialized, the \dacquirer can use $\msg.\textsf{value}$ to deposit the initial funds. Additionally, during the registration phase, the \dacquirer can continue to increase funding to attract more \dproviders to participate in training. After each increase in funding, the smart contract also emits an \textsf{EpochDeposit} event to notify all \dproviders. Therefore, when \dacquirer invokes the $\deposit$ function, the smart contract checks whether it is still in the registration phase. Once the registration phase is completed, the \dacquirer will no longer be able to deposit funds.

\parab{\textbf{Commit Models.}} The $\register$ function allows \dproviders to commit their local models on the smart contract, which means they participate in the current epoch of training. Upon receiving notification of the \textsf{NewEpoch} event, \dproviders can commit local models by $\register$ function. During the model registration phase, \dproviders need to upload the hash and signature of their model in order to prevent malicious \dproviders from uploading incorrect models in subsequent model transactions, which ensures the integrity and security of data transactions. 
When the registration phase has ended, the $\register$ function sets the $\train[\epoch].isReg$ flag to $\tru$, indicating that the round of model registration has ended and unregistered \dproviders cannot participate in the current epoch of training.

\parab{Prepare Aggregation.} In the preparation phase, the \dacquirer needs to accomplish two goals through the smart contract. The first is to record the amount of reward allocated to each participant. The allocation ratio needs to be consistent with the aggregation weight of each local model in order to ensure the success of verification. The second goal is to select the verification parameters for random sampling through the $\vdf$ function. The $\vdf$ function based on smart contracts ensures that the generation of random numbers is not influenced by the \dacquirer or \dproviders. After achieving these goals, the smart contract sets $\train[\epoch]$.isPrepared to $\tru$.


\parab{Commit Public Inputs.} After the preparation phase, the \dacquirer should commit the verification key $vk$, the quantized public input $\mathbb{X}^{t,c}$, and the proof $\pi^t_{\textsf{agg}}$ to blockchain by the \commit \xspace function for later verification. 

\parab{Verify Aggregation.}Verifying the integrity of aggregation is the main goal of the smart contract. Because the $\vrf$ function requires a large amount of gas fee, the smart contract ensures that each epoch only needs to be validated successfully once. 
The core function of the verification phase is $\verify$, which implements the zk-SNARKs verification of the aggregation process. To reduce the gas fees required for contract deployment, we use the verifying key parameter as the input value for contract execution. We have designed the $\cvk$ function, which can reconstruct the verifying key by the input parameters. Otherwise, in order to perform zk-SNARKs verification within the block gas limit, we use the precompiled contracts in Ethereum for elliptic curve pairing operations, such as the $\add$, $\smul$, $\negate$, etc. After the successful execution of $\vrf$, the smart contract broadcasts \textsf{EpochVerified} event to all the \dproviders to claim their reward.

\parab{Claim Reward.} After the aggregation is verified, \dproviders can withdraw their reward from the smart contract account through the $\claim$ function at any time. The smart contract will check whether the address of the \dprovider belongs to a participant in the training round to prevent malicious \dproviders from claiming funds. Also, after the smart contract sends the relevant return to the \dproviders, the participant's proportion in the smart contract is set to $0$. This means that if a \dprovider attempts to withdraw his reward again, they will only consume the gas fee and receive no additional reward.

\subsection{The CNN Architecture of the Global Model Used for CIFAR Dataset}
\label{subsec:cnn_arch}

\begin{table}[]
\begin{tabular}{@{}cc@{}}
\toprule
Layer                           & Size                                   \\ \midrule
Input                           & $3 \times 32 \times 32$                \\
CNN + ReLU, Max Pooling & $3 \times 15 \times 3$, $2 \times 2$   \\
CNN + ReLU, Max Pooling & $15 \times 75 \times 4$, $2 \times 2$  \\
CNN + ReLU, Max Pooling & $75 \times 375 \times 3$, $2 \times 2$ \\
Linear + ReLU                   & $1500 \times 400$                      \\
Linear + ReLU                   & $400 \times 120$                       \\
Linear + ReLU                   & $120 \times 84$                        \\
Linear                          & $84 \times 10$                         \\ \bottomrule
\end{tabular}
\caption{The CNN architecture of the global model used for CIFAR~\cite{krizhevsky2009learning} dataset.}\label{tab:cnncifar}
\end{table}

Table \ref{tab:cnncifar} shows the architecture of the Convolutional Neural Network (CNN) used for the CIFAR~\cite{krizhevsky2009learning} dataset in our experiments.

The CNN has seven layers. The first layer is the input layer, which takes in images that are $32 \times 32$ pixels with three channels. The next three layers are CNN layers, each followed by a ReLU activation function and a max pooling layer. The first CNN layer has 15 filters of size $3 \times 3$, the second CNN layer has 75 filters of size $4 \times 4$, and the third CNN layer has 375 filters of size $3 \times 3$. The size of kernels in all max pooling layers is $2 \times 2$. The last four layers are fully connected layers, each followed by a ReLU activation function, except for the last one. The size of fully connected layers are $1500 \times 400$, $400 \times 120$, $120 \times 84$, and $84 \times 10$, respectively. The output of the last layer is a vector of 10 numbers, which represent the probability of the input image belonging to each of the 10 classes. The class with the highest probability is the predicted class of the input image.